\documentclass[%
 aip,
 amsmath,amssymb,
 reprint,%
]{revtex4-1}

\usepackage{graphicx}
\usepackage{dcolumn}
\usepackage{bm}

\usepackage[utf8]{inputenc}
\usepackage[T1]{fontenc}
\usepackage{mathptmx}

\begin{document}

\preprint{AIP/123-QED}

\title[Point Defects in Crystals of Charged Colloids]{Point Defects in Crystals of Charged Colloids}

\author{Rinske M. Alkemade}
\affiliation
{Soft Condensed Matter, Debye Institute of Nanomaterials Science, Utrecht University}
\author{Marjolein de Jager}
\affiliation
{Soft Condensed Matter, Debye Institute of Nanomaterials Science, Utrecht University}
\author{Berend van der Meer}
\affiliation
{Department of Chemistry, Physical and Theoretical Chemistry Laboratory,University of Oxford, South Park Road, Oxford OX1 3QZ, United Kingdom}
\author{Frank Smallenburg}
\affiliation
{Universit\'e Paris-Saclay, CNRS, Laboratoire de Physique des Solides, 91405 Orsay, France}
\author{Laura Filion}
\email[]{l.c.filion@uu.nl}
\affiliation
{Soft Condensed Matter, Debye Institute of Nanomaterials Science, Utrecht University}

\date{\today}

\begin{abstract}
 Charged colloidal particles -- both on the nano and micron scales --  have been instrumental in enhancing our understanding of both atomic and colloidal crystals. These systems can be straightforwardly realized in the lab, and tuned to self-assemble into body-centered cubic (BCC) and face-centered cubic (FCC) crystals. While these crystals will always exhibit a finite number of point defects, including vacancies and interstitials – which can dramatically impact their material properties -- their existence is usually ignored in scientific studies. Here, we use computer simulations and free-energy calculations to characterize vacancies and interstitials in both FCC and BCC crystals of point-Yukawa particles. We show that, in the BCC phase, defects are surprisingly more common than in the FCC phase, and the interstitials manifest as so-called crowdions: an exotic one-dimensional defect proposed to exist in atomic BCC crystals. Our results open the door to directly observing these elusive defects in the lab.
\end{abstract}

\maketitle

\section{Introduction}

Suspensions of charged colloids are among the most fundamental systems in colloidal science. These systems, consisting of charged colloidal spheres suspended in a solvent containing salt, which screens the Coulombic repulsions between the spheres, have been extensively studied using experiments, simulations, and theory \cite{alexander1984charge, kremer1986phase, robbins1988phase, monovoukas1989experimental, sirota1989complete, hamaguchi1997triple, hynninen2003phase, yethiraj2003colloidal, hsu2005charge, royall2006re, el2011measuring, smallenburg2011phase, kanai2015crystallization, arai2017surface, chaudhuri2017triple}. In the case of single-component, spherical colloids, the bulk phase behavior is extremely well understood, with impressive quantitative comparisons between theory and experiment \cite{monovoukas1989experimental, sirota1989complete, kanai2015crystallization}. These quantitative comparisons have been facilitated by the highly tunable nature of experimental systems of charged colloids \cite{yethiraj2003colloidal, van2013sterically, kodger2015precise}.  From these studies we know that for sufficiently high densities or strongly charged particles,  identically charged colloids self-assemble into one of two crystal structures, depending on the degree of screening. Broadly, for low salt concentrations, where the screening is weak, the system forms a body-centered-cubic (BCC) crystal, while high salt concentrations result in a face-centered-cubic (FCC) crystal. 

In equilibrium, such crystalline phases always feature a finite concentration of defects. These defects, like vacancies and  interstitials, can have a profound impact on the mechanical, optical, and electronic properties of crystalline materials. In the realm of colloid science, where the creation of new materials to manipulate light is one of the overarching goals, the presence of defects strongly affects optical properties \cite{yan2005incorporation, rengarajan2005effect, nelson2011epitaxial}. It is therefore perhaps surprising, that despite the massive body of literature on crystals formed by charged colloids, little is known about how defects manifest in their 3d crystalline phases.

In three dimensions, some of the earliest work on defects in colloidal crystals focused on point defects (vacancies and interstitials) in single-component hard-sphere crystals \cite{bennett1971studies, pronk2001point,pronk2004large}. This colloidal model system forms an FCC crystal, with relatively few point defects in equilibrium: at melting the crystal is predicted to have approximately $10^{-4}$ vacancies and $10^{-8}$ interstitials per lattice site. Subsequent studies have explored {\it e.g.} the local structural impact of defects \cite{lin2016measuring, van2017diffusion, vansaders2018strain}, the diffusion of vacancies and interstitials \cite{bennett1971persistence, van2017diffusion}, and the emergence of stacking faults \cite{hoogenboom2002stacking,pronk1999can,marechal2011stacking,pusey1989structure} in hard-sphere systems.
However, in general,  defects in 3d colloidal crystals have received relatively little attention due, at least in part, to the expectation that they do not occur in large quantities in equilibrium.  

A notable exception is the relatively recent prediction that simple cubic crystals of repulsive particles frequently exhibit large numbers ($\approx 0.06$ per lattice site) of vacancies that are spread over a row of lattice sites in one dimension \cite{van2017phase,van2018revealing,smallenburg2012vacancy}.
These 1d vacancies, predicted for simple cubic crystals, are reminiscent of so-called interstitial crowdions. This intriguing type of interstitial defect was proposed by Paneth in 1950 \cite{paneth1950mechanism}, to explain anomalous self-diffusion in BCC crystals of alkali metals. In this picture, the defect is expected to spread out over multiple lattice sites arranged along a one-dimensional line, resulting in preferential diffusion along that direction. In the atomic realm, explorations of these defects has been largely focused on simulations \cite{derlet2007multiscale, nguyen2006self, osetsky2003one, han2002self,zepeda2004strongly},  and simple theoretical models used to capture their behaviour \cite{kontorova1938theory,landau1993model, kovalev1993theoretical, braun1998nonlinear, dudarev2003coherent, fitzgerald2008peierls}.

To date, an analogue to these defects in a colloidal realization of a BCC crystal -- which would allow for direct observation in real time using {\it e.g.} confocal microscopy -- is lacking. It is therefore intriguing to explore how interstitials manifest in colloidal BCC crystals, and in particular in systems that can be directly, and even quantitatively, reproduced in an experimental setting.

Here we use computer simulations to explore both the concentration and structure of point defects in both  BCC and FCC crystals of one of the most fundamental models for screened charged particles --  the point Yukawa model.  Our results predict that this fundamental system forms a direct colloidal realization of crowdion interstitials in BCC crystals. Moreover, we find that BCC exhibits significantly higher concentrations of point defects, and hence expect that these crowdions play an important role in controlling the material properties of the crystal.
Importantly, given the substantial concentration of crowdions predicted to occur in equilibrium, our results pave the way to directly observing these rare and elusive defects in colloidal experiments.

\section{Model}

We consider a system of $N$ charged colloids of diameter $\sigma$ suspended in a solvent containing ions characterized by  an inverse Debye screening length $\kappa_D$ and Bjerrum length $\lambda_B$.  Within Derjaguin-Landau-Verwey-Overbeek (DLVO) theory, the effective potential between the colloids is given by 
\begin{align}
\beta \phi(r) =\frac{\epsilon}{r} e^{-\kappa_D r},
\label{yukawa}
\end{align}
where
\begin{equation}
    \epsilon = \frac{Z^2  \lambda_B e^{\kappa_D \sigma}}{\left(1+\kappa \sigma / 2\right)^2}
\end{equation}
with $Z$ the charge of the colloids in electron charges, and $\beta=1/k_B T$, with $k_B$ the Boltzmann constant and $T$ the temperature. Note that this so-called Yukawa potential describes, not only charged colloids, but also has been widely applied in the study of dusty plasmas \cite{ivlev2012complex}.

\newcommand{\widthpd}{0.8\linewidth}
\begin{figure}
\begin{tabular}{ll}
	a) & \\[-0.3cm]
	\includegraphics[width=\widthpd]{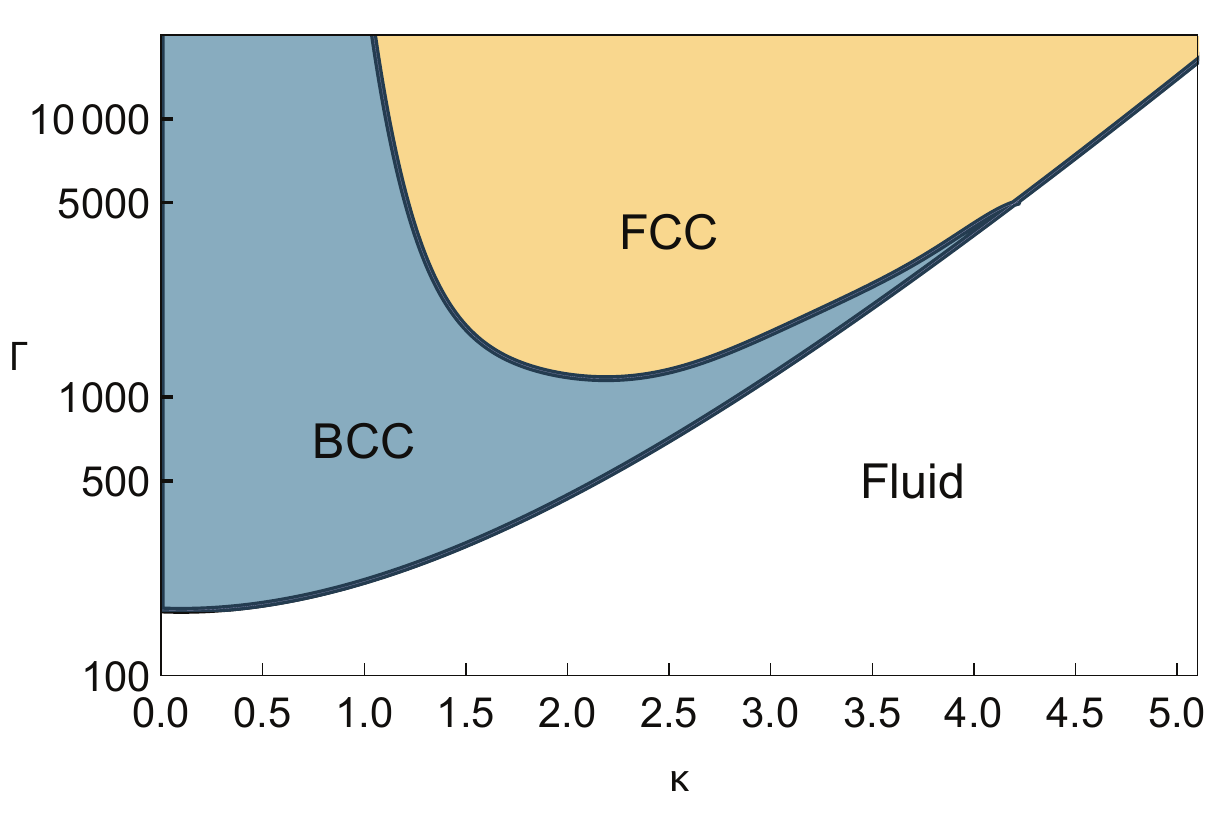}\\
	b) & \\[-0.3cm]
	\includegraphics[width=\widthpd]{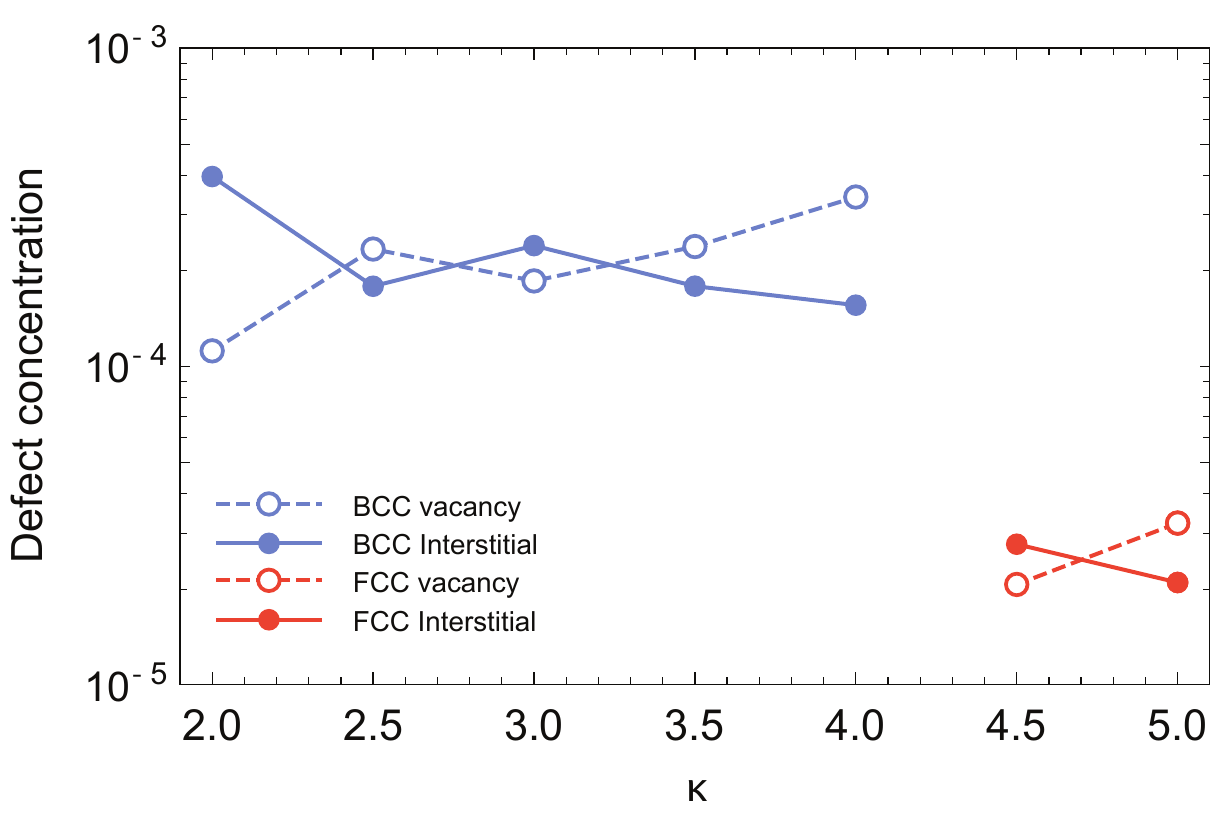}\\
	c) & \\[-0.3cm]
	\includegraphics[width=\widthpd]{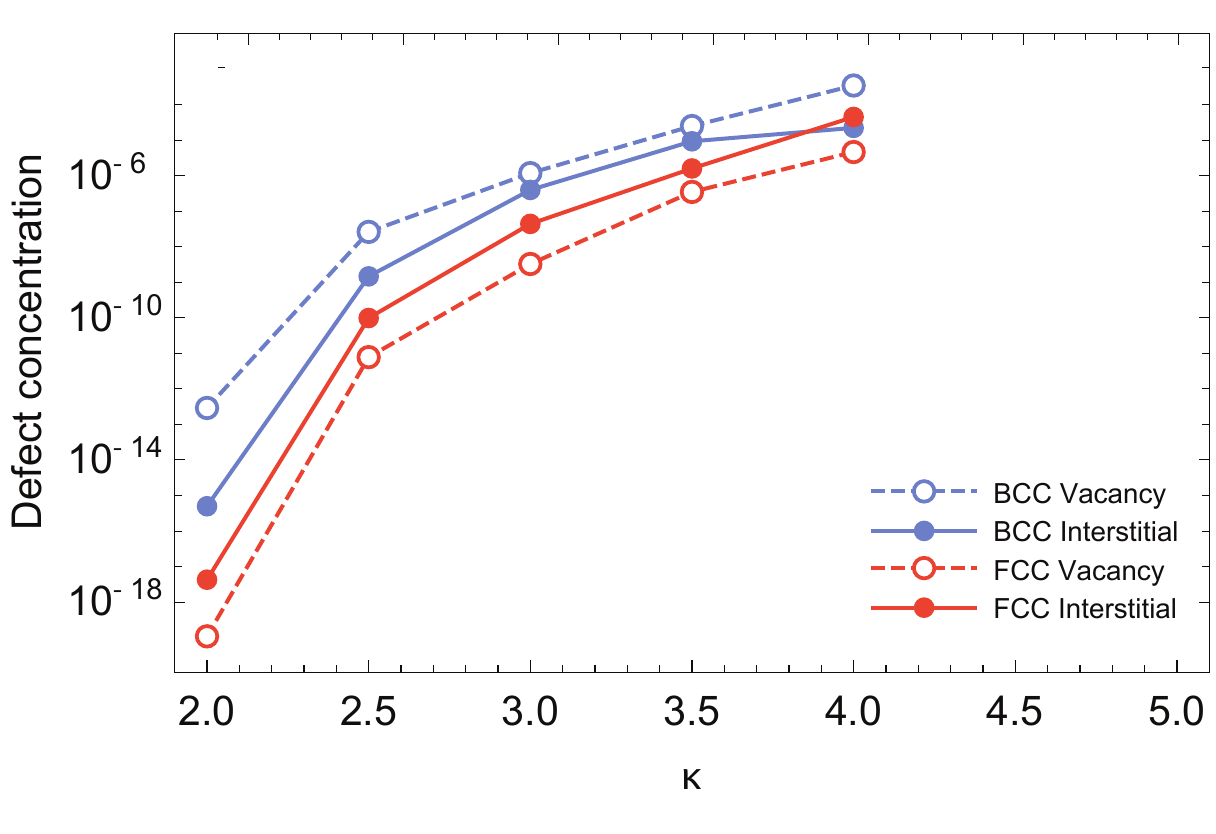}
	\end{tabular}
	\caption[width=1\linewidth]{a) Phase diagram of Yukawa systems in the ($\kappa, \Gamma$) plane, with the phase boundaries from Refs. \onlinecite{hamaguchi1997triple, hynninen2003phase}. Vacancy and interstitial concentrations plotted along b) fluid-crystal phase boundary  and c) BCC-FCC phase boundary. We estimate an error of up to 0.2$k_B T$ in our calculation of the $\mu^{\mathrm{vac(int)}}$, leading to an approximate error of a factor of 1.2 in the concentrations. }
	\label{fig:phasediagram}
\end{figure}

Conveniently, the phase behaviour of this system can be fully characterized by two dimensionless parameters, namely
\begin{align*}
\Gamma = \frac{\epsilon}{a k_B T}, \\ \kappa = a\kappa_D,
\end{align*}
with $a=\left(\frac{4\pi N}{3V}\right)^{-\frac{1}{3}}$ the Wigner Seitz radius. 

The phase diagram for this system has been explored extensively using theory, simulations, and experiments. In Fig. \ref{fig:phasediagram}, we show the phase behavior, using the phase boundaries approximated in Refs. \onlinecite{hamaguchi1997triple, hynninen2003phase}.  It consists of a fluid phase and two crystal phases: face-centered cubic (FCC) and body-centered cubic (BCC), with all phase boundaries corresponding to first-order phase transitions. Note that the coexistence regions here are  all small and have been simply presented as lines, similar to Refs. \onlinecite{hamaguchi1997triple,hynninen2003phase}. 
In this paper we will explore the behaviour of point defects associated with the crystals that appear in this 2d phase diagram.

 \section{Methods}
 
 \subsection{General Simulation Details}
We used Monte Carlo simulations in the $NVT$-ensemble with periodic boundary conditions \cite{frenkel2001understanding}, where the particles interact {\it via} the point Yukawa potential (Eq. \ref{yukawa}). The potential was truncated and shifted such that the shift was never more than  $10^{-5} k_B T$. The system size was chosen to always be large enough to accommodate this choice, such that the cutoff range is less than half the box length. 

The system sizes were chosen depending on the phase in question. For studying defect concentrations we used systems containing between 250-1500 particles. For select points of the smallest systems we examined whether doubling the system size mattered, and in all cases it had no discernible effect on the defects concentrations.  For studying the shape of the defects, we used system ranging from 1000-3500 particles and again ensured that the system size was not affecting the results.

\subsection{\label{Concentration}Concentration of defects}

To determine the vacancy concentration, we make the assumption that the defect concentration is sufficiently low  that i) the defects are not interacting, ii) the equation of state of the crystal is unaffected by the presence of defects.  In this case, the free energy of a system of $N$ particles in a volume $V$ at temperature $T$, can be written 
\begin{eqnarray}
\beta F^{\text{vac}}(N,V,T)&=& \beta Mf^{\text{df}}(N,V,T)+ \beta (M-N) f^{\text{vac}} \nonumber \\&&  + N\log \frac{N}{M} + (M-N)\log\frac{M-N}{M},
\label{eq:freeenergy}
\end{eqnarray} 
where $M > N$ is the number of lattice sites,  $f^\mathrm{df}$ is the free energy per particle of the defect-free crystal, and $f^{\text{vac}}$ is the free energy associated with creating a single vacancy at a specific lattice site. 

Taking the Legendre transform to turn this Helmholtz free energy into a Gibbs free energy, and minimizing with respect to the number of lattice sites $M$, we find that the equilibrium concentration of vacancies is given by
 \begin{equation}
\left\langle n_{\text{vac}}\right\rangle \equiv \left\langle\frac{M-N}{N}\right\rangle =  \exp[-\beta \mu^{\text{vac}}],
\label{eq:vaccon}
\end{equation}
where $\mu^{\text{vac}}$ is defined as $ f^{\text{vac}}(\rho_M,T)+ \mu^{\text{df}}(P,T)$ with $\mu^{\text{df}}(P,T)$ the chemical potential of a defect-free crystal and $\rho_M = M/V$ the density of lattice sites. Note that $P$ is the pressure.

For a crystal containing interstitials, where $M<N$, we use a similar approach yielding an equilibrium concentration of interstitials 
\begin{equation}
\left\langle n_{\text{int}}\right\rangle \equiv \left\langle\frac{N-M}{N}\right\rangle= \exp[-\beta \mu^{\text{int}}],
\label{eq:intcon}
\end{equation} 
\noindent with  $\mu^{\text{int}}=f^{\text{int}}(\rho_M,T)-\mu^{\text{df}}(P,T)$. Here $f^{\text{int}}(\rho_M,T)$ is the free energy associated with creating an interstitial at a specific lattice site. 
In order to obtain the concentration of point defects for various points along the phase boundary of the Yukawa crystal, we thus need to measure $f^{\text{vac}}(\rho_M,T)$, $f^{\text{int}}(\rho_M,T)$ and $\mu^{\text{df}}(P,T)$ in a Yukawa crystal. Because it is not possible to measure these free energies directly in a Monte Carlo simulation, we use thermodynamic integration as described below.

To obtain the chemical potential $\mu^{\mathrm{df}}$ of the defect-free crystals, we first use the Frenkel-Ladd method \cite{frenkel1984new} to obtain the Helmholtz free energy. We then combine this with the pressure, measured via the virial expression \cite{frenkel2001understanding}, to determine the chemical potential.

We now turn our attention to the method for finding the free energies $f^{\text{int}}(\rho_M,T)$ and $f^{\text{vac}}(\rho_M,T)$, associated with creating point defects. 
In the case of a vacancy, we break $f^{\text{vac}}$ up into two contributions: $f^{\text{vac}} = f^\text{shrink} + f^\text{remove}$, where $f^\text{shrink}$ is associated with turning one of the particles of a defect-free crystal into a non-interacting particle, and $f^\text{remove}$ is associated with removing this non-interacting particle. Similarly, in the case of an interstitial we first compute the free energy, $f^\text{add}$, associated with inserting a non-interacting particle and then calculate the free energy, $f^\text{grow}$, associated with turning this non-interacting particle into a normal-interacting particle. Note that in all cases, the particle associated with a defect is confined to a single Wigner-Seitz cell. 

We then calculate the total free energy for the interstitial using $f^\mathrm{int} = f^\mathrm{grow} + f^\mathrm{add}$.
The free energies associated with $f^\text{add}$ and $f^\text{remove}$ are given by:
\begin{equation}
f^{\text{add}} =  -k_BT\ln\left(\frac{V^{WS}}{\Lambda^3}\right).
\label{idealgasE2}
\end{equation}
and 
 \begin{equation}
f^{\text{remove}} = k_BT\ln\left(\frac{V^{WS}}{\Lambda^3}\right),
\label{idealgasE}
\end{equation}
 where $V^{WS}$ is the volume of the Wigner-Seitz cell and $\Lambda$ is the thermal DeBroglie wavelength.
 
To calculate $f^\mathrm{shrink}$, we use thermodynamic integration with an auxiliary Hamiltonian 
\begin{equation}
U_\lambda = (1-\lambda)U_{\text{0}} + \lambda U_\text{non-int},
\label{eq:auxmethod1}
\end{equation}
with $U_{\text{0}}$ the normal interaction potential of our system, and $U_{\text{non-int}}$ the potential energy of a system where one particle is non-interacting. Following standard thermodynamic integration, the free-energy difference between a crystal with one non-interacting particle and a defect-free crystal is then given by,
\begin{equation}
f^{\text{shrink}}=F^{\text{non-int}}-F^{\text{df}}=\int_{0}^{1}d\lambda\left\langle U_{\text{non-int}}-U_{\text{0}} \right\rangle_{\lambda},
\label{eq:thermointvac}
\end{equation}
with $F^{\text{non-int}}$ the Helmholtz free energy of a crystal containing one non-interacting particle. The ensemble average, $\left\langle... \right\rangle_{\lambda}$ is evaluated using the auxiliary potential given in Eq. \eqref{eq:auxmethod1}. The free energy $f^\mathrm{grow}$ is determined following the same method. Note that in both cases, we evaluate the integral numerically using 34 different values of $\lambda$.

However, while in theory this method works fine, in practice the sampling can become very slow. When the system is at $\lambda=0$ we are evaluating the energy difference in Eq. \ref{eq:thermointvac} using the potential $U_{\lambda =0} =U_{\text{non-int}} $. This means that without any energy penalty, the non-interacting particle can come very close to other particles if those particles are near their Wigner-Seitz cell boundary. Because we compute the potential energy difference with the system where our particle does have interactions, the term $\left\langle U_{\text{non-int}}-U_{\text{0}} \right\rangle_{\lambda =0}$ can become very large - in the interstitial case even infinitely large. Due to these large energy differences, the simulation needs a long time to get a reliable answer for the value $\left\langle U_{\text{non-int}}-U_{\text{0}} \right\rangle_{\lambda =0}$.  

To circumvent this problem we alter the potential. Instead of letting it diverge at $r=0$ as it normally would do, we assume that the potential increases linearly below a certain defined value $r_{\text{alter}}$. By doing so the potential has a finite value, $U_\text{max}$, at $r=0$.  If we evaluate $\left\langle U_{\text{non-int}}-U_{\text{0}} \right\rangle_{\lambda =0}$ for higher $\lambda$'s this altered potential will not have any influence as the particles will never have a distance \textit{r} with  $r<r_{\text{alter}}$ due to the energy penalty. However for small $\lambda$, we avoid the large energies. Because in the end we integrate over the energy difference (see Equation \eqref{eq:thermointvac}), this alteration to the potential has no influence on the final free energy. We find that the precise values for  $U_\text{max}$ and $r_{\text{alter}}$ do not matter, as long as we make sure $r_{\text{alter}}$ is sufficiently small that for higher values of $\lambda$, \textit{r} will almost never be smaller than $r_{\text{alter}}$. We checked this by running the same simulation twice for  different values of $U_\text{max}$ and $r_{\text{alter}}$.


\newcommand{\figwidth}{0.28\linewidth}
\newcommand{\figwidthB}{0.27\linewidth}
\newcommand{\legendsize}{0.08\linewidth}

\section{Results}

\begin{figure*}
    \centering
    \begin{tabular}{lll}
        & \,\,\,\, a) & \,\,\,\, b)\\[-0.3cm]
        \includegraphics[width=\legendsize,trim= 0cm 0.2cm 0.0cm 0.0cm]{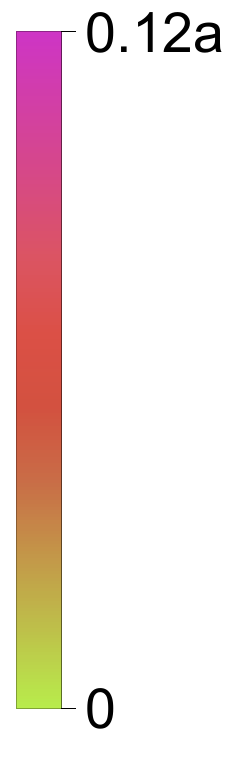} & 
        \hspace{0.7cm}
        \includegraphics[width=\figwidth]{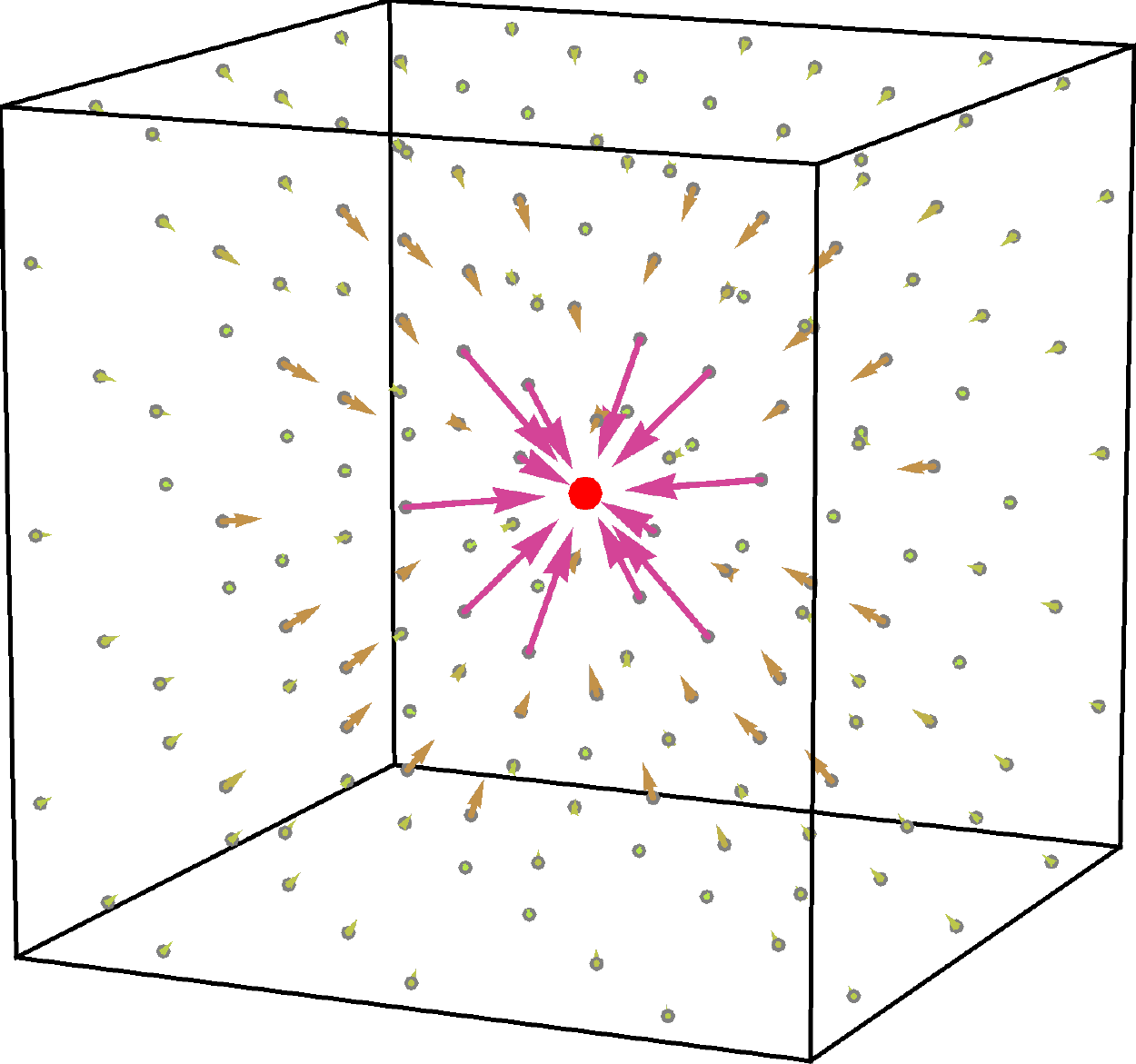} & 
        \hspace{0.7cm}
        \includegraphics[width=\figwidthB]{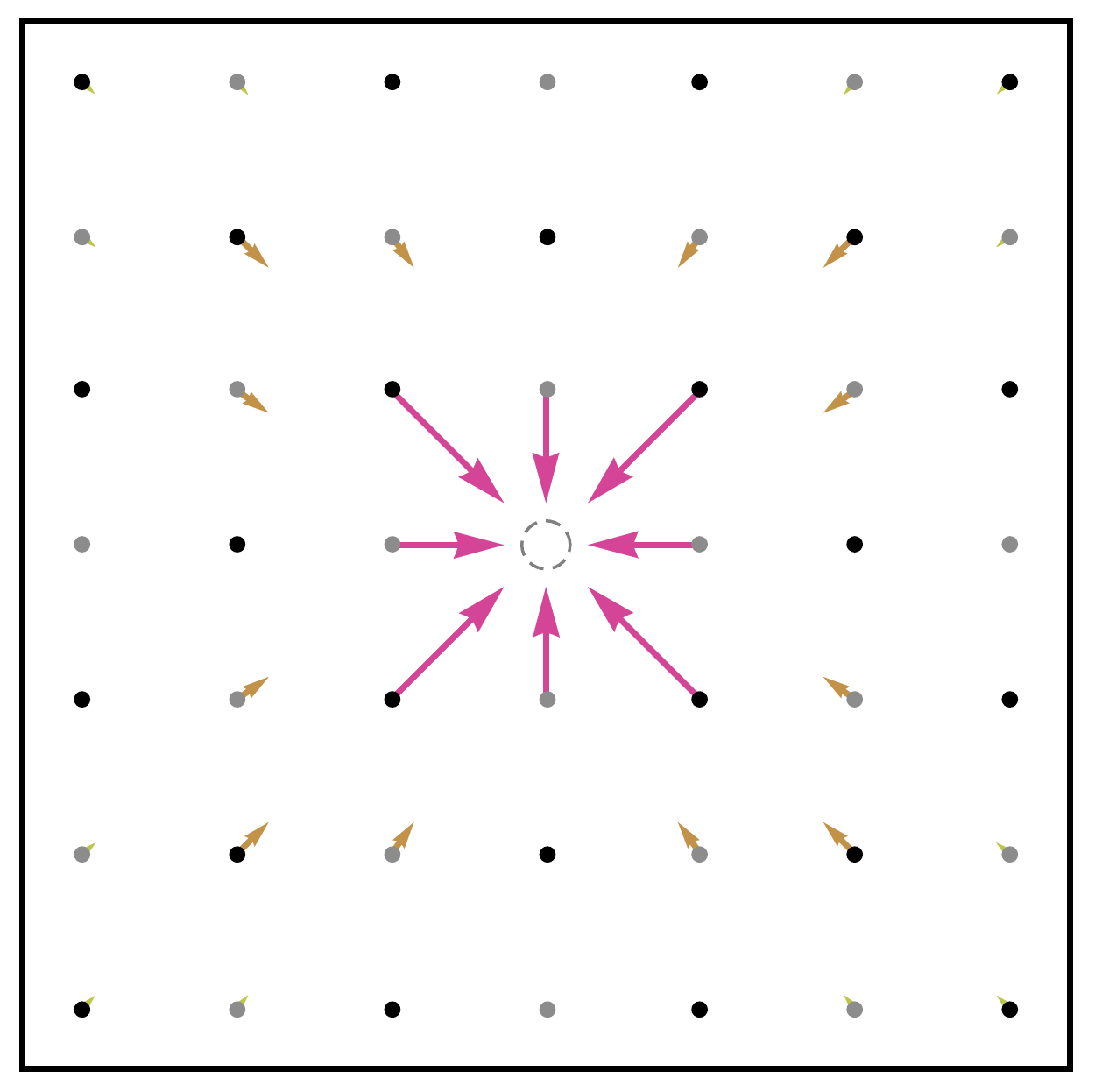} \\
        & \,\,\,\, c) & \,\,\,\, d)\\[-0.3cm]
         \includegraphics[width=\legendsize,trim= 0cm 0.2cm 0.0cm 0.0cm]{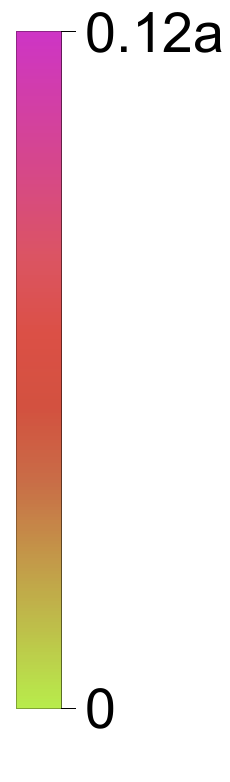} & 
         \hspace{0.7cm}
         \includegraphics[width=\figwidth]{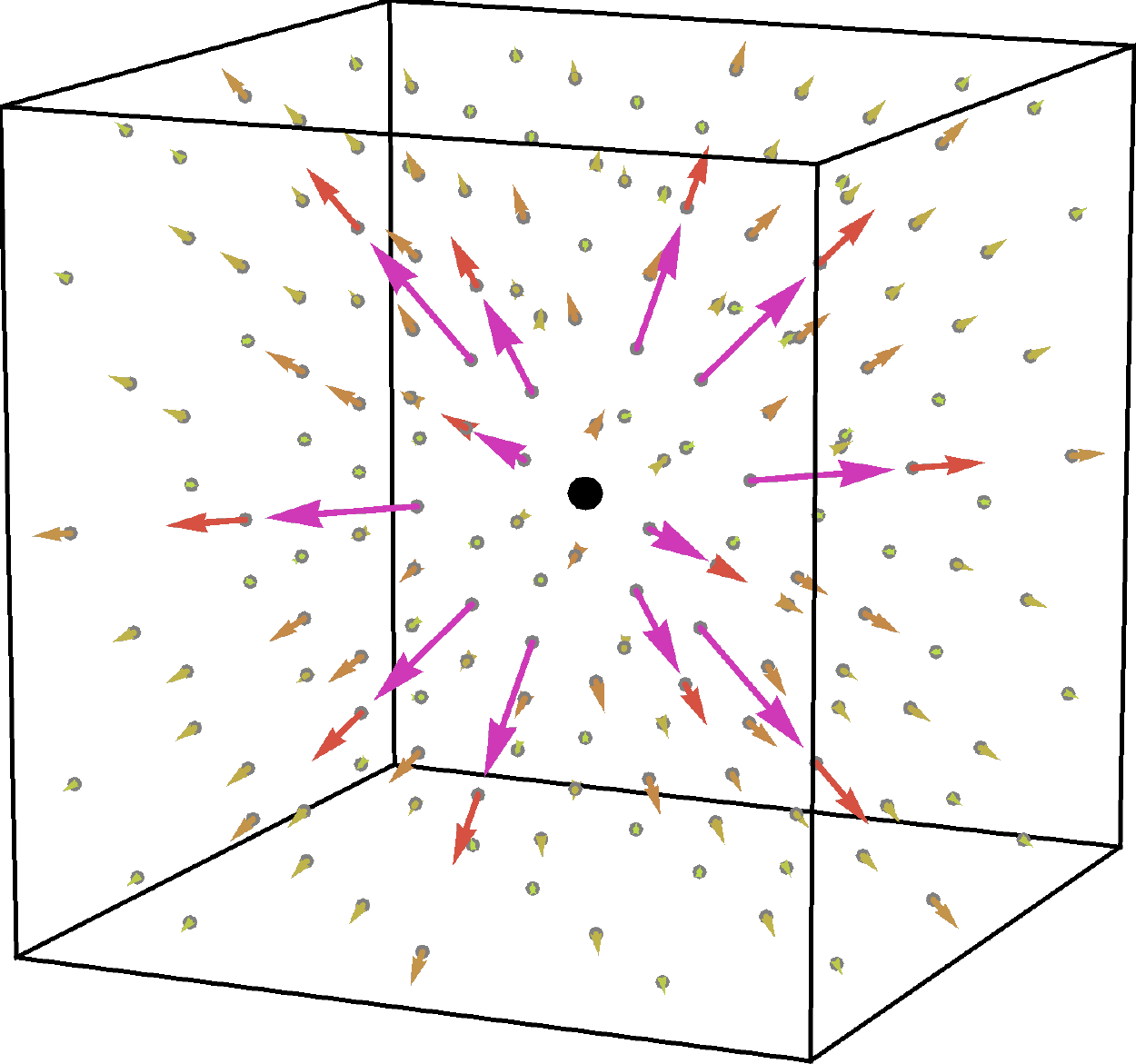} & 
         \hspace{0.7cm}
         \includegraphics[width=\figwidthB]{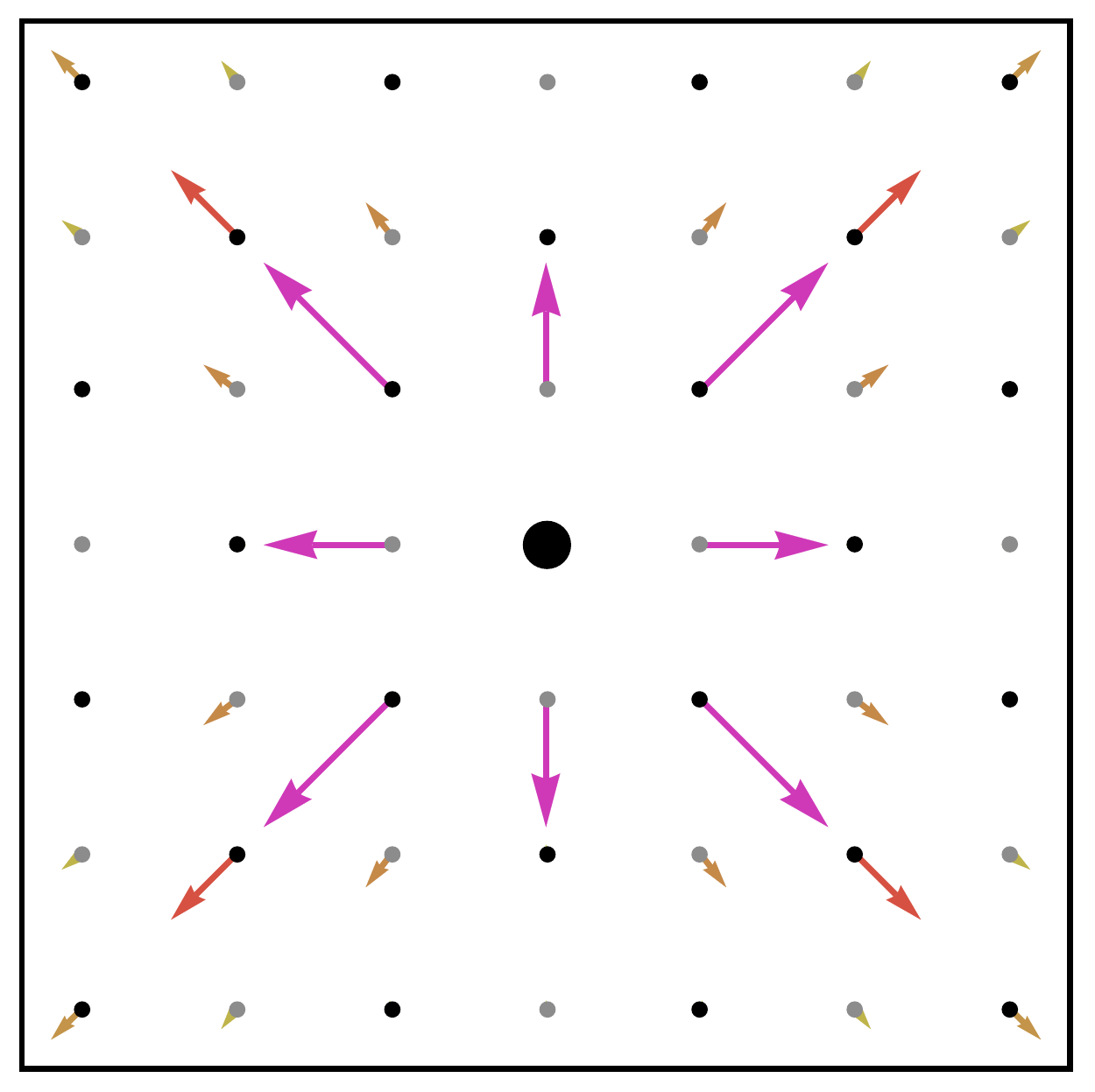} \\
        & \,\,\,\, e) & \,\,\,\, f)\\[-0.3cm]
        \includegraphics[width=\legendsize,trim= 0cm 0.5cm 0.0cm 0.0cm]{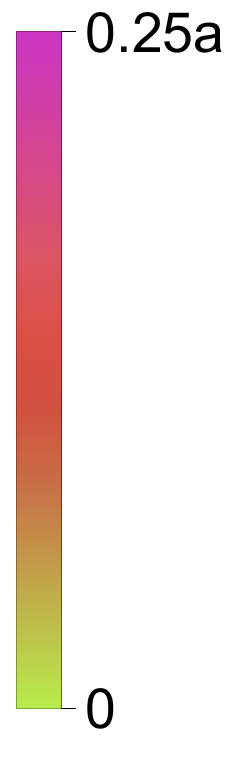} & 
        \hspace{0.7cm}
        \includegraphics[width=\figwidth]{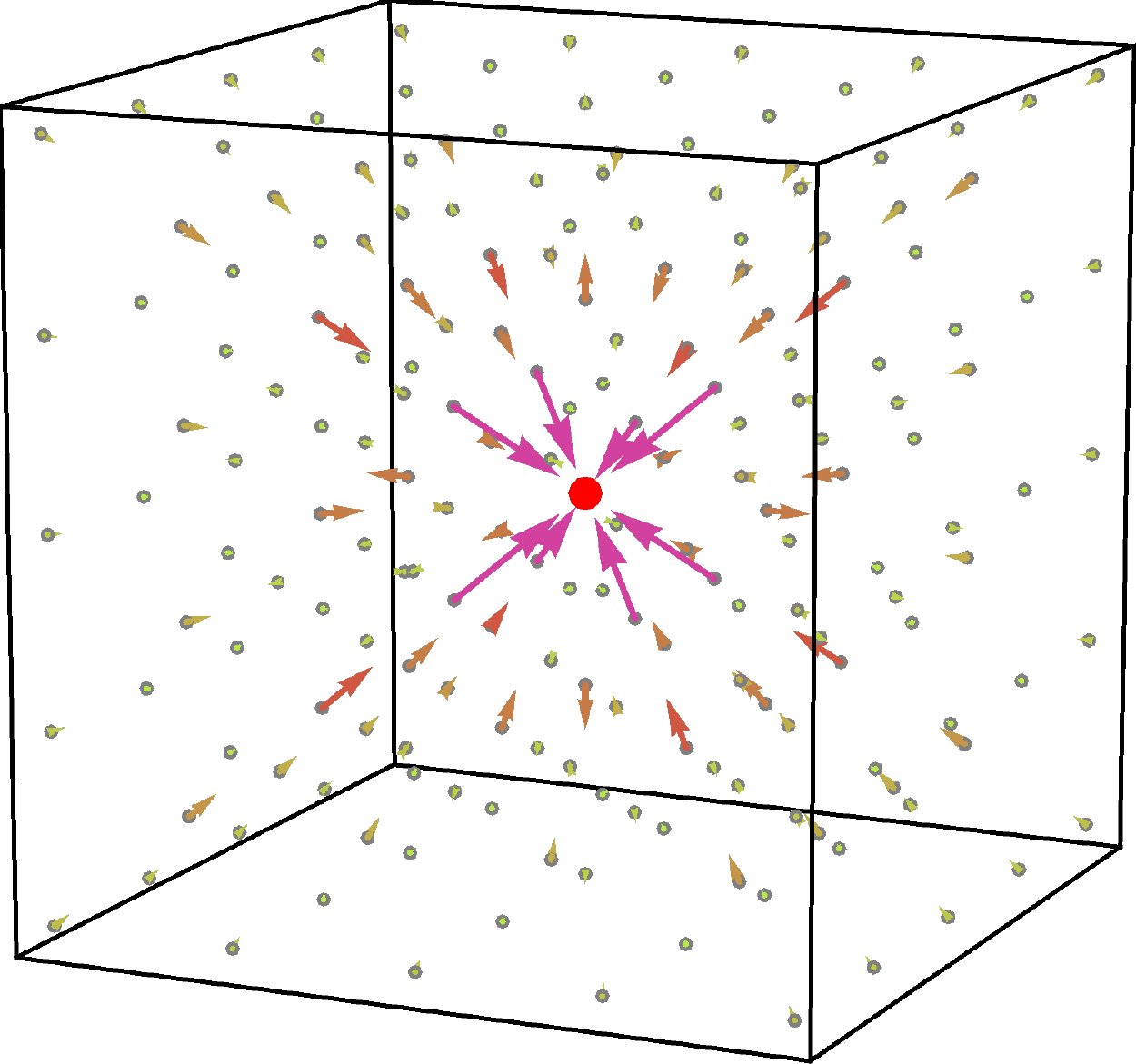} & 
        \hspace{0.7cm}
        \includegraphics[width=\figwidthB]{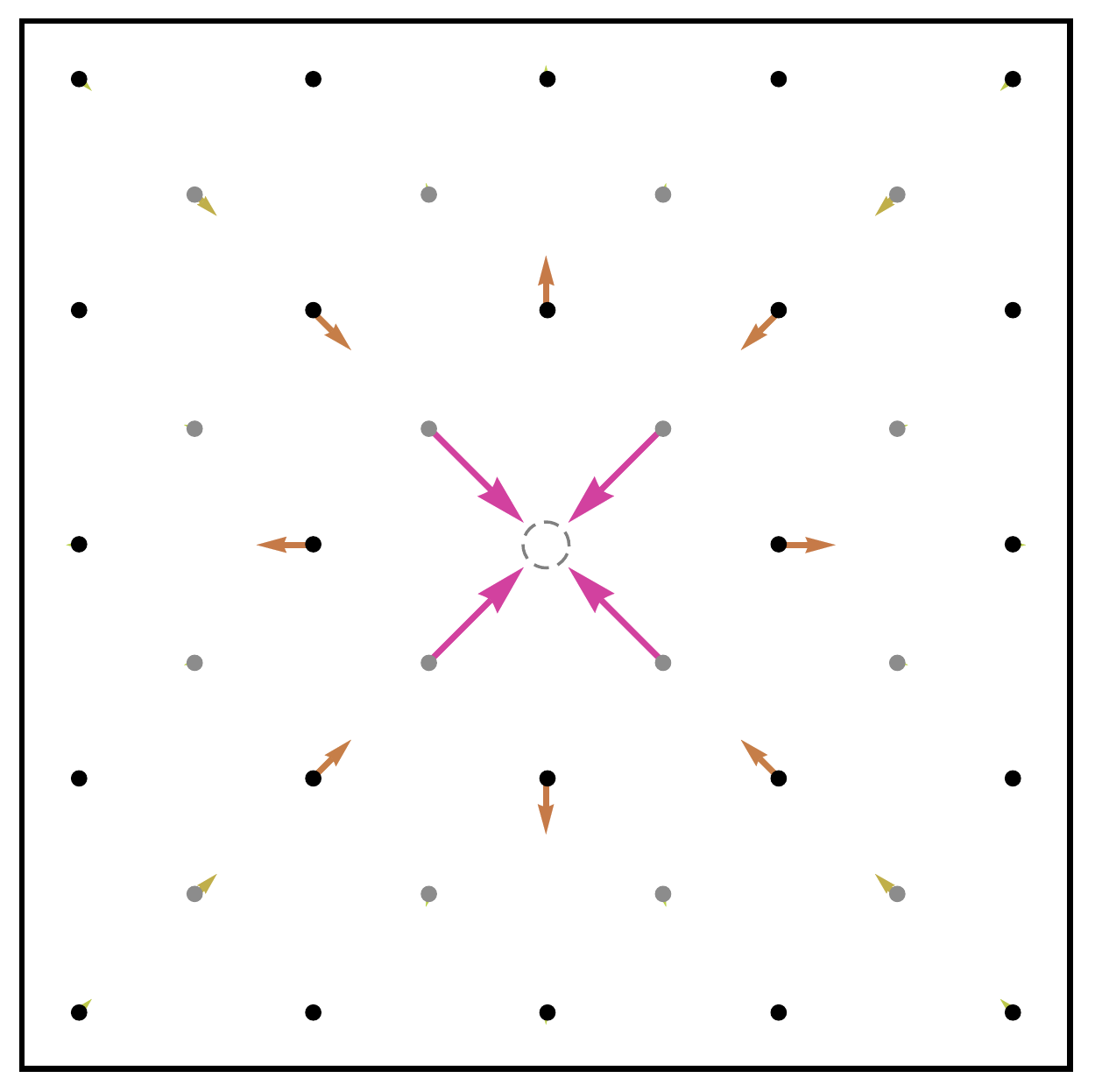} \\
        & \,\,\,\, g) & \,\,\,\, h)\\[-0.3cm]
         \includegraphics[width=\legendsize,trim= 0cm 0.5cm 0.0cm 0.0cm]{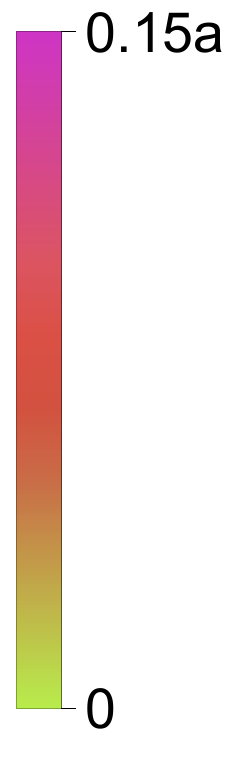} & 
         \hspace{0.7cm}
         \includegraphics[width=\figwidth]{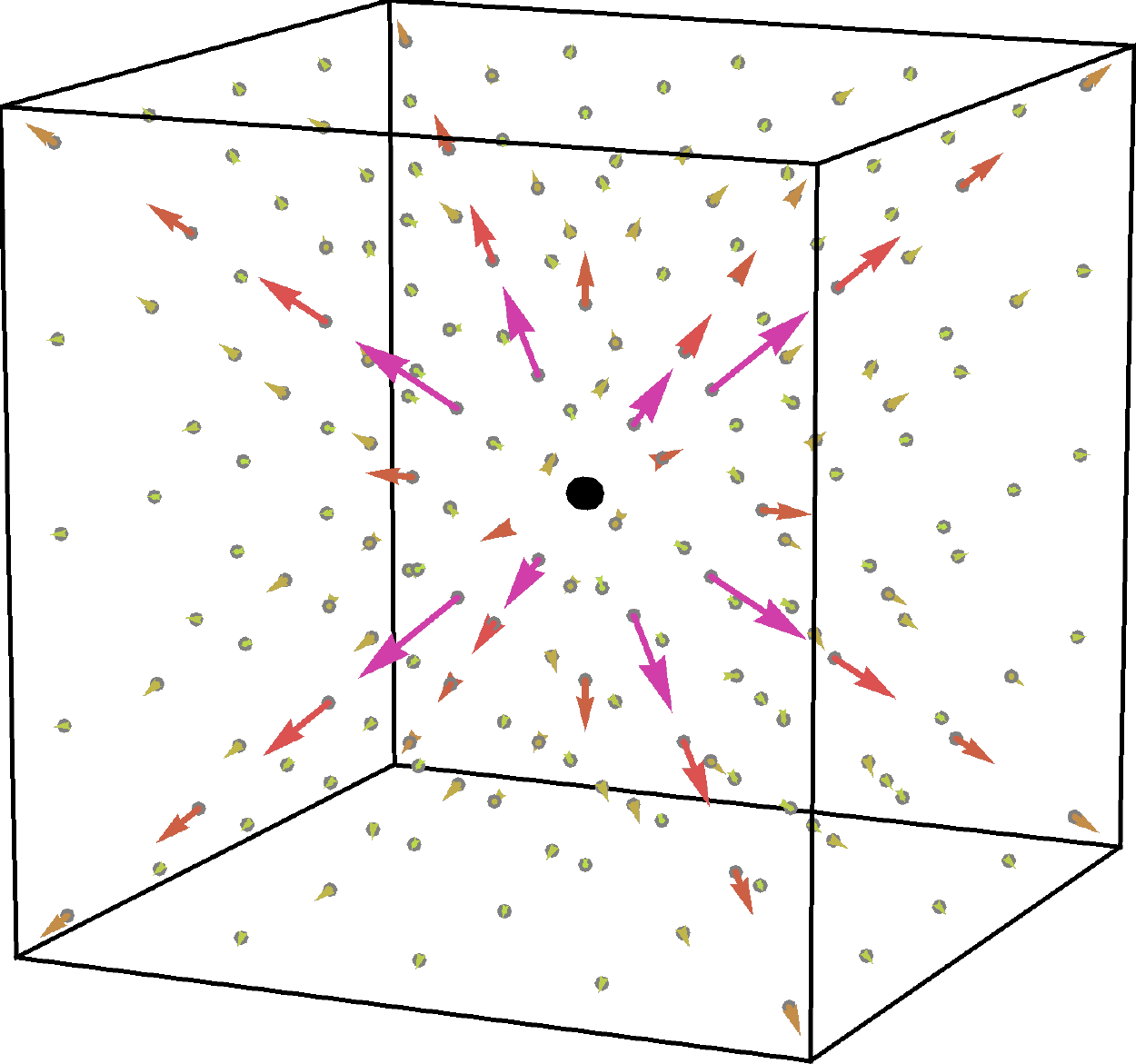} & 
         \hspace{0.7cm}
         \includegraphics[width=\figwidthB]{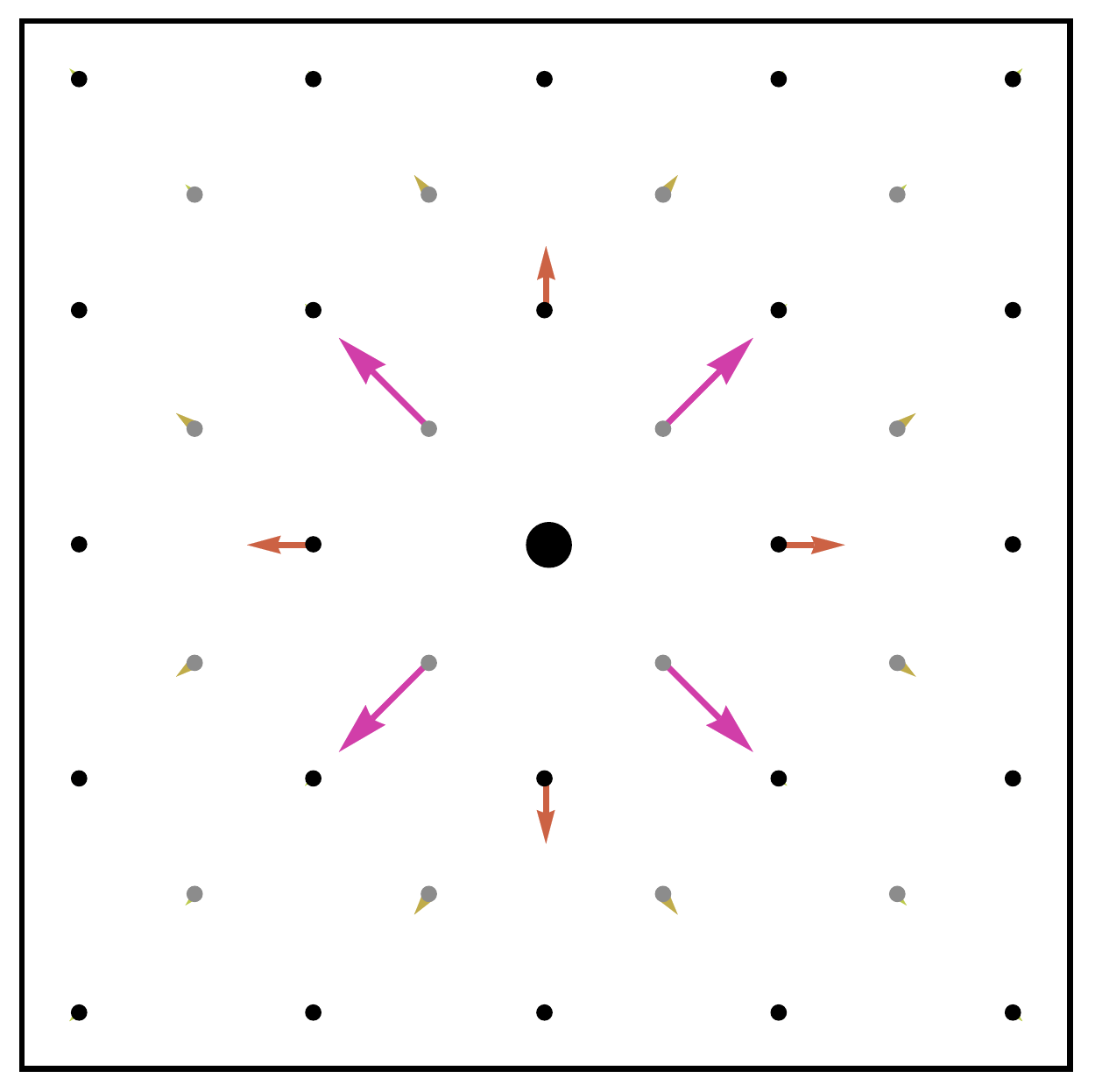} 
    \end{tabular}
    \caption{Average lattice deformation due to a point defect in a-d) the FCC crystal at $\kappa=3.5$ and $\Gamma=2565$ and e-h) the BCC crystal at $\kappa=3.5$ and $\Gamma=2400$. 
    a-b,e-f) Deformation due to a vacancy (indicated by red or a dotted sphere) and c-d,g-h) deformation due to an interstitial (indicated by a black sphere).
    Left: 3d representation of part of the simulation box. The gray points represent the lattice sites. Right: projection of the displacement vectors on two (100) planes on top of each other. The black points represent the lattice sites that lie in the plane of the defect and the gray points the lattice sites in the neighboring plane.
    In all figures the size of the arrows is exaggerated, but the color of the arrows indicates the deformation in terms of the Wigner Seitz radius $a$.
    }
    \label{fig:defectstructure}
\end{figure*}

\newcommand{\FKfigs}{\linewidth}
\begin{figure*}
    
    \begin{tabular}{lll}
    &a)&\\[-0.3cm]
    \includegraphics[width=\legendsize,trim= 0cm -1.5cm 0.0cm 0.0cm]{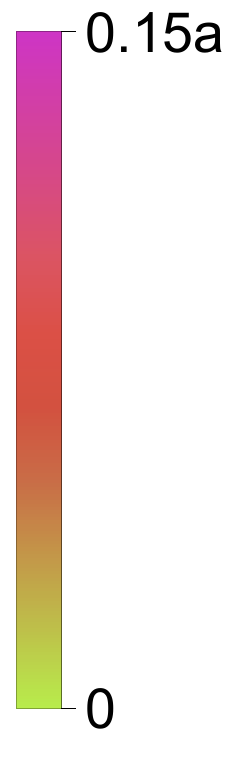} & \includegraphics[width=0.42\linewidth]{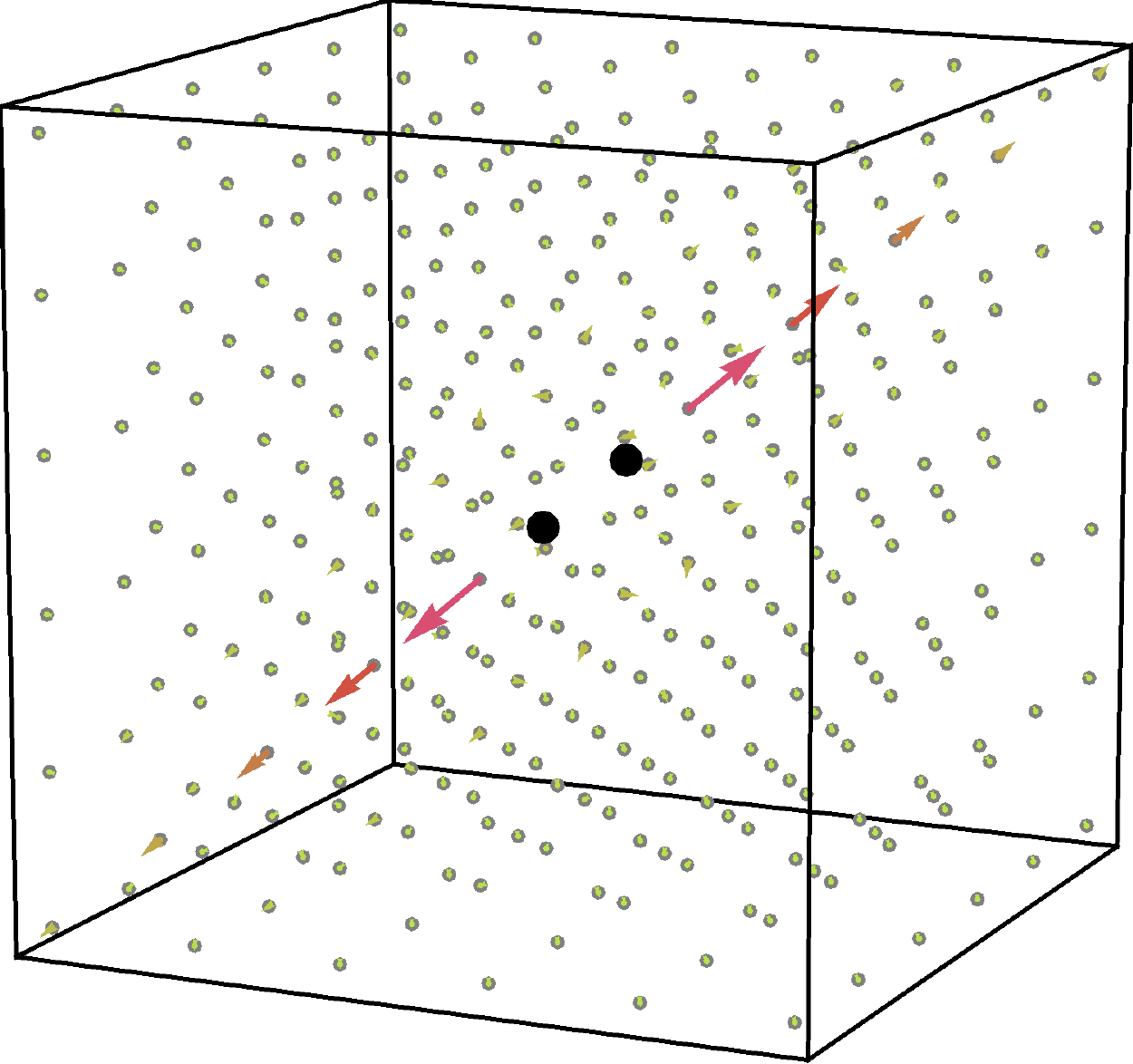} &
    \begin{minipage}{6cm}
    \begin{raggedright}
    
    \vspace{-5.75cm}
    b)\\[-0.3cm]
    \includegraphics[width=\FKfigs]{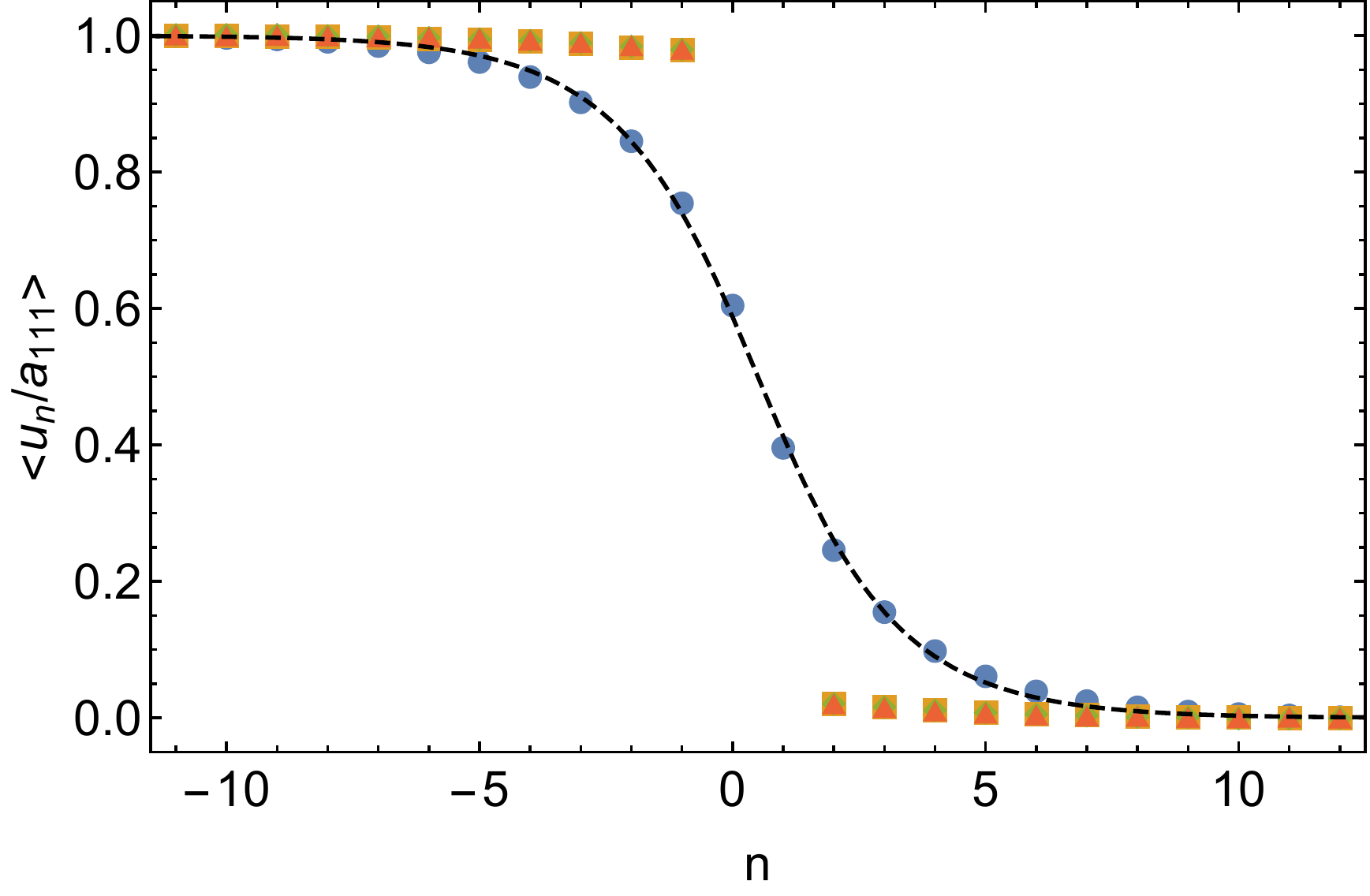}  \\
    c)\\[-0.3cm]
    \includegraphics[width=\FKfigs]{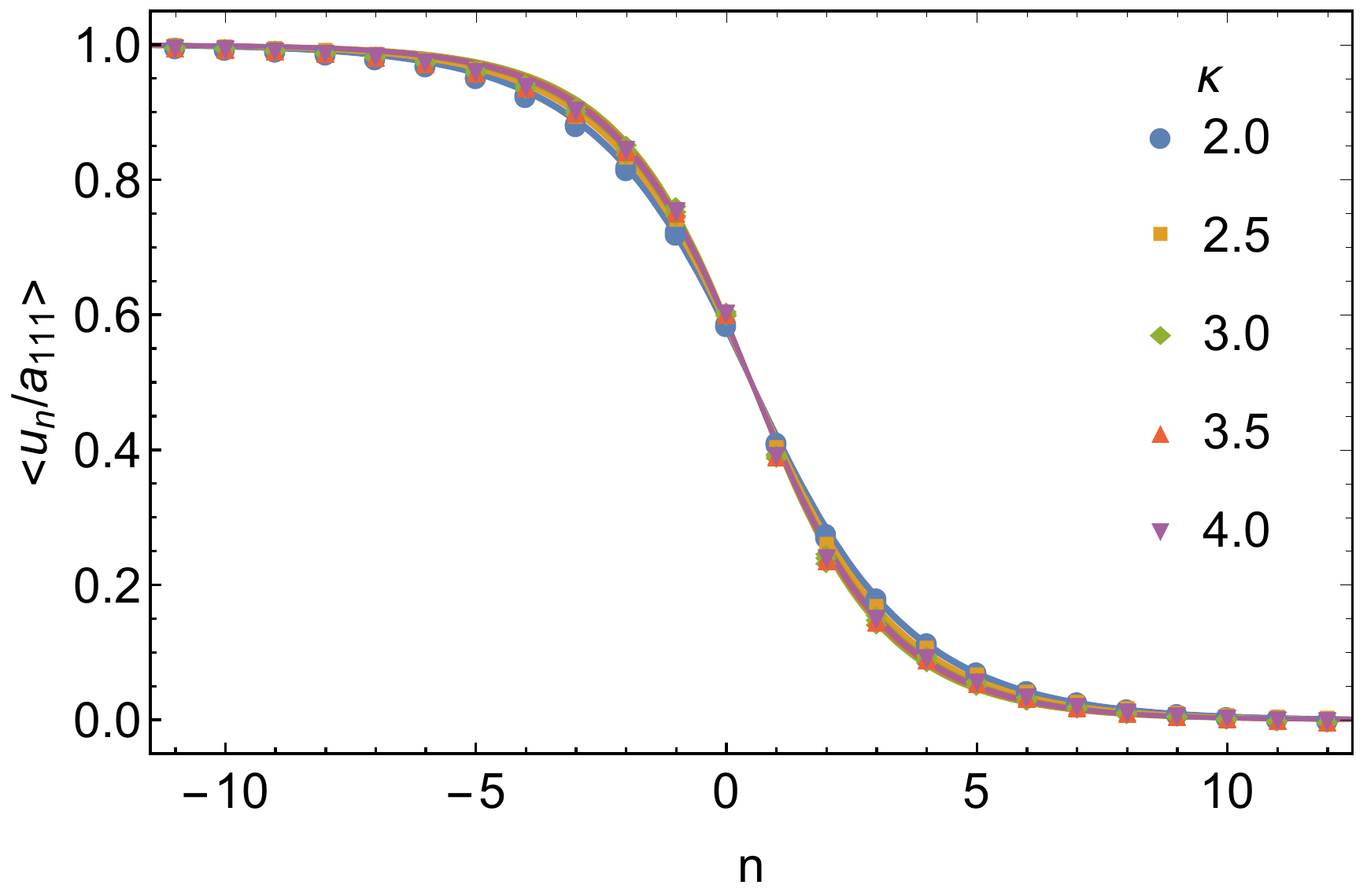}
    \end{raggedright}
    
    \end{minipage}
        \end{tabular}
    \caption{a) Lattice deformation due to an interstitial in the BCC crystal at $\kappa=3.5$ and $\Gamma=2400$. The gray points represent the lattice sites of part of the simulation box and the black spheres represent the positions of the interstitial and its companion. 
    The size of the arrows is exaggerated, but the color of the arrows indicates the deformation in terms of the Wigner Seitz radius $a$. Note that these deformations are averaged over multiple configurations which have been rotated so that the defect always points along the same direction.
    b) Displacement $u_n$ along the four $\langle 111 \rangle$ directions for the same system as a). The blue dots indicate $u_n$ along the direction of the crowdion and the dashed line represents the corresponding fitted soliton solution. 
    c) $u_n$ along the direction of the crowdion for $\kappa=2.0$ and $\Gamma=1000,1100,1205$ (blue dots), $\kappa=2.5$ and $\Gamma=1000,1100,1205$ (yellow squares), $\kappa=3.0$ and $\Gamma=1205,1400,1600$ (green diamonds), $\kappa=3.5$ and $\Gamma=2200,2300,2400$ (orange triangles), $\kappa=4.0$ and $\Gamma=4000,4100,4200$ (purple triangles). The lines represent the corresponding fitted soliton solutions.
    }
    \label{fig:crowdion}
\end{figure*}

We start our investigation by exploring the equilibrium concentration of  vacancies and interstitials in both crystals. As a starting point we focus on state points in the vicinity of the fluid-crystal phase boundary -- the region on the phase diagram that is expected to have the highest concentration of defects. To predict these concentrations we make the assumption that the defects do not interact, and that their effect on the pressure of the system is negligible. We can then use a combination of Monte Carlo simulations and thermodynamic integration  to extract the defect concentrations.  The result is shown in Fig. \ref{fig:phasediagram}b \footnote{Note that we have taken the phase boundaries from Ref. \onlinecite{hamaguchi1997triple}, and have ensured that the crystal phases do not melt in our simulations.}.

Clearly, along the fluid-crystal line, BCC appears to have more defects of both types than FCC.  More specifically, for BCC both the vacancy and interstitial concentrations are on the order of $10^{-4}$, while for FCC the concentrations are closer to $10^{-5}$. For vacancies, these concentrations are similar to those found for hard spheres at coexistence ($10^{-4}$) \cite{pronk2001point}. However, the interstitial concentration in both cases is orders of magnitude higher than the $10^{-8}$ predicted for interstitials in hard-sphere crystals at the fluid-crystal phase boundary \cite{pronk2001point}.

To make a more direct comparison of the behaviour of the two crystals, we then calculated the defect concentrations along the FCC-BCC phase line. From Fig. \ref{fig:phasediagram}c we observe again that BCC generally has more defects than FCC; while the difference in the interstitial concentration is small, the difference in vacancy concentration varies from two to eight orders of magnitude. Clearly, BCC generally exhibits more equilibrium point defects than FCC.  

To explore the large differences we observe between FCC and BCC, we now turn our attention to the structure of the point defects in these two crystals. To determine the structure, we performed $NVT$ MC simulations with a single vacancy or interstitial present.  To prevent the defect from hopping during our analysis, we confined all particles to their Wigner-Seitz cells \cite{van2017phase, van2020high} and then measured the average location of each particle during the simulation. 

The results for a vacancy in FCC are shown in Fig. \ref{fig:defectstructure} a-b,  while e-f depict the average deformation associated with a vacancy in a BCC crystal. In both crystals, as expected, the largest deformation is associated with the first shell of neighbours --  however, it is approximately twice as large in the case of the BCC crystal (note the different scaling on the color bars).  This likely arises due to the lower number of particles in the first shell of BCC, 8 in comparison to the 12 in FCC. In the BCC crystal, the particles are less strongly caged by their neighbors, and hence more free to move into the space opened up by the vacancy. More interesting is the behaviour of the second shell of neighbours.  While in FCC all particles again deviate in the direction of the vacancy, in BCC half the particles move towards the vacancy while the other half move away from the defect. The end result is a much larger change in energy of the crystal upon removing the particle: in the system shown in Fig. \ref{fig:defectstructure}, the average difference is  $\beta\Delta U_\text{FCC} = -35.6$  and $\beta\Delta U_\text{BCC} = -42.9$ in the FCC and BCC crystals, respectively. This difference contributes directly to the difference in free-energy cost for creating a vacancy in the two crystals. In short, the BCC structure is better able to take advantage of the vacancy to reduce its local potential energy, which helps to alleviate the cost of creating a vacancy and hence makes them more prevalent.

The interstitials turn out to be an even more interesting case. In Fig. \ref{fig:defectstructure} c-d, and g-h we plot the average displacement of particles from their lattice sites in FCC and BCC crystals, respectively. In comparison  to the vacancy case, these defects appear at first rather similar: in both crystals the average deviations due to 
the interstitial are mainly along the lines pointing along the  nearest-neighbour directions, and decay slowly through a number of neighbouring shells. In the FCC crystal this means that particles lying  along the six $\left< 110 \right>$ lines are displaced the most, similar to what was found in hard-sphere crystals \cite{van2017diffusion}. In the BCC crystals it is the particles lying along the four $\left< 111 \right>$ lines.

\begin{figure*}
    \begin{tabular}{llll}
        \hspace{-0.5cm} a) &  & \hspace{1.1cm} b) &  \\ [-0.25cm]
        \includegraphics[height=0.28\linewidth,trim= 0cm 0.2cm 0.0cm 0.0cm]{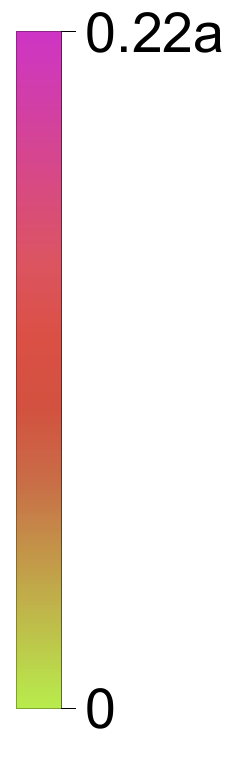} & 
        \hspace{-0.7cm}
        \includegraphics[width=\figwidth]{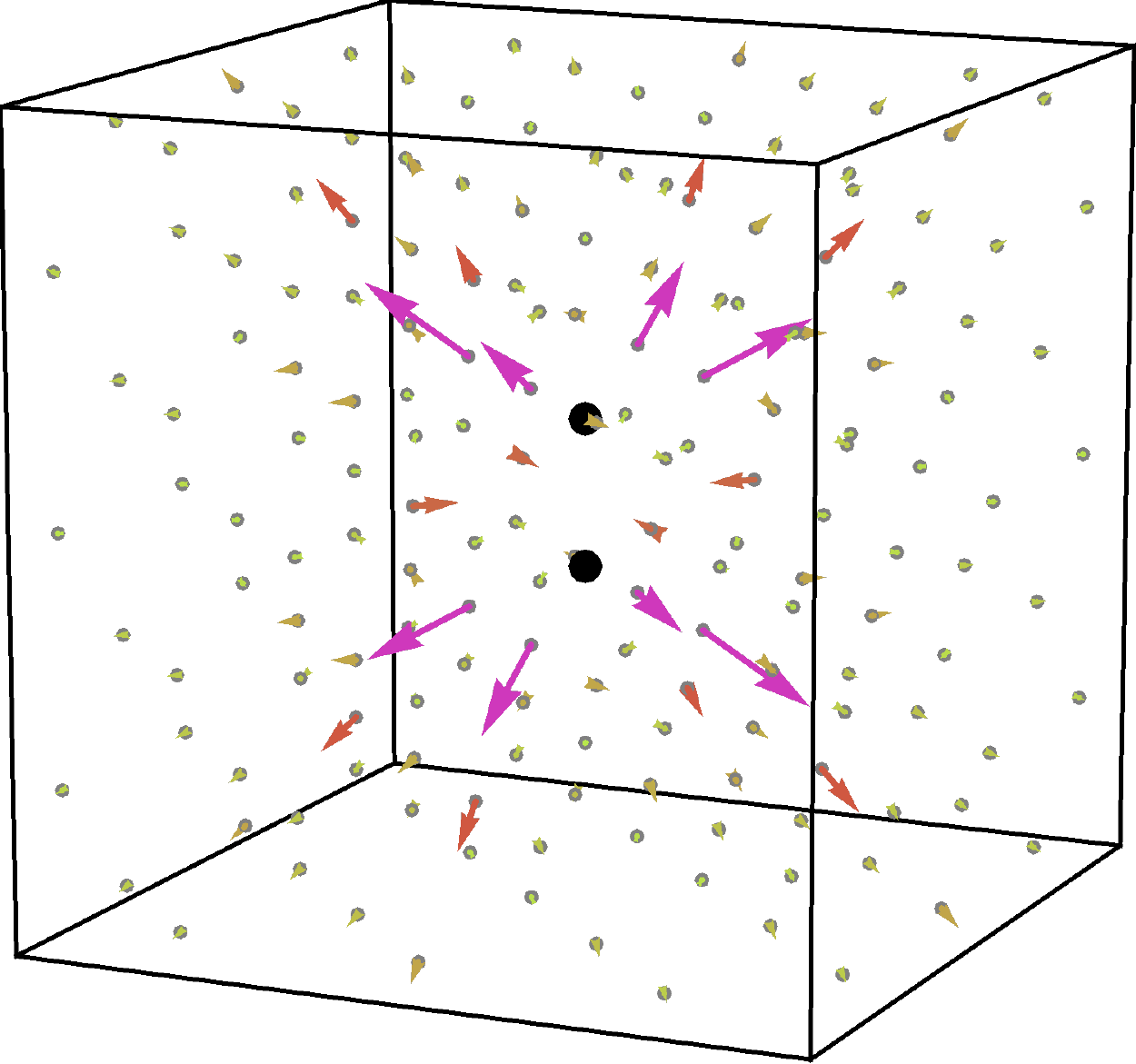} & 
        \hspace{1.4cm}
         \includegraphics[height=0.28\linewidth,trim= 0cm 0.2cm 0.0cm 0.0cm]{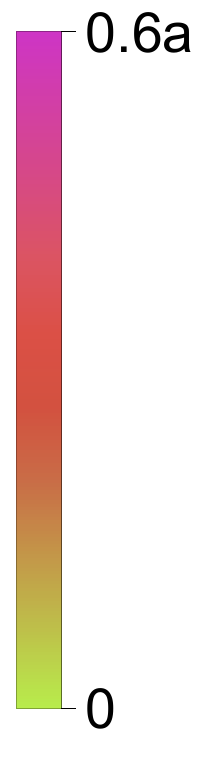} & 
        \hspace{-0.7cm}
         \includegraphics[width=\figwidth]{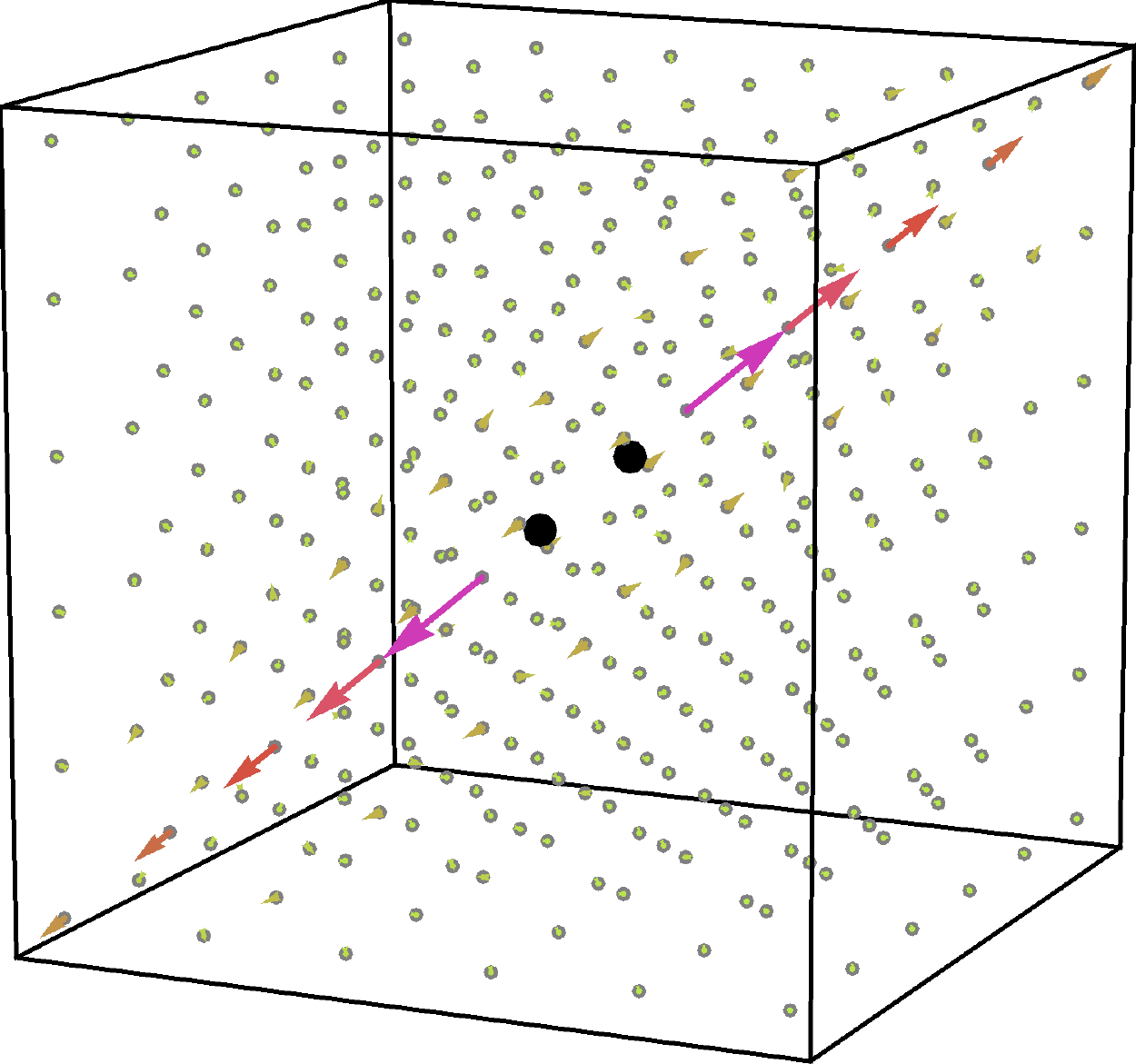} 
         \hspace{0.7cm}
    \end{tabular}
    
    \vspace{0.7cm}
    \begin{tabular}{ll}
       \hspace{.45cm} c) & \hspace{0.8cm} d)  \\ [-0.25cm]
        \hspace{0.5cm} \includegraphics[width=0.4\textwidth]{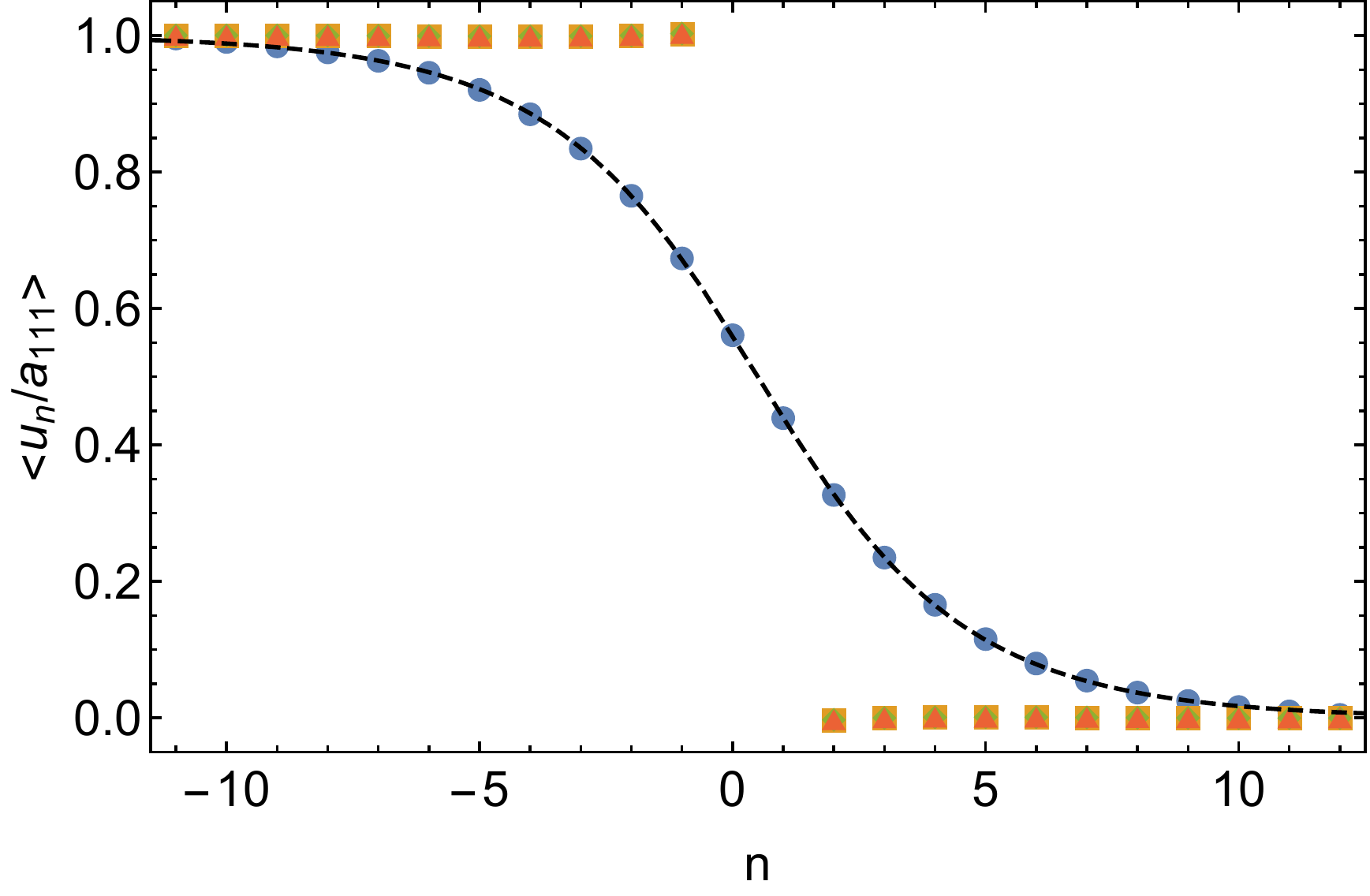} & 
    \hspace{1cm}  \includegraphics[width=0.4\textwidth]{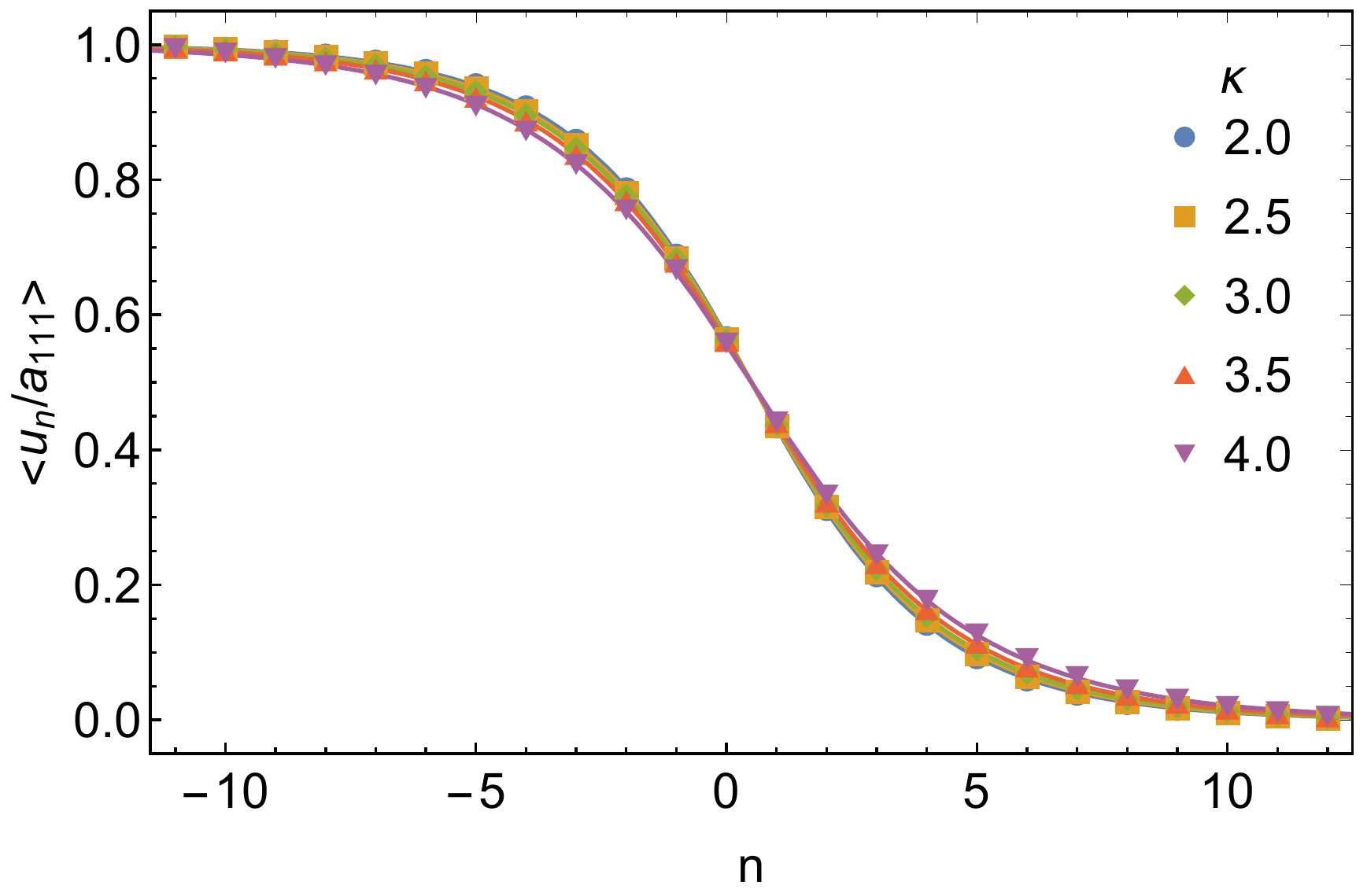}          
    \end{tabular}
    \caption{Quenched lattice deformation due to an interstitial in a) the FCC crystal and b) the BCC crystal, both at $\kappa=3.5$.
    The gray points represent the lattice sites of part of the simulation box and the black spheres represent the actual positions of the interstitial and its companion. The size of the arrows is exaggerated, but the color of the arrows indicates the deformation in terms of the Wigner-Seitz radius $a$.
    c) Displacement $u_n$ along the four $\left< 111 \right>$ directions for the same system as b). The blue dots indicate $u_n$ along the direction of the crowdion and the dashed line represents the corresponding fitted soliton solution.
    d) $u_n$ along the direction of the crowdion for five different $\kappa$. The lines represent the corresponding fitted soliton solutions. 
    }
    \label{fig:defectstructure2}
\end{figure*}

Interestingly, however, the averaged displacements do not tell the full story. If we examine the instantaneous realization of an interstitial, there is a spontaneous symmetry breaking in the displacement of neighboring particles along different directions, especially in the case of BCC. One clear way of demonstrating this is to quench a system containing an interstitial to high values of $\Gamma$, such that the system minimizes its potential energy. As shown in Fig. \ref{fig:defectstructure2}, this quench has a particularly remarkable effect on the BCC crystal. While in the case of FCC (Fig. \ref{fig:defectstructure2}a), the defect takes on a normal dumbbell structure with a 3d displacement field around it, in BCC the defect becomes one-dimensional (Fig. \ref{fig:defectstructure2}b):  only particles along one of the four $\left< 111 \right>$ lines are displaced significantly. With this knowledge in mind, we can recalculate the average displacement field for an interstitial by first rotating each configuration so that the defect is always oriented along the same axis. The result is shown in Fig. \ref{fig:defectstructure2} for the quenched system and in  Fig. \ref{fig:crowdion} for the system at finite $\Gamma$.

This one-dimensional configuration strongly resembles a so-called  crowdion: an exotic 1d defect proposed to exist in some metallic BCC crystals \cite{paneth1950mechanism}. In order to characterize the structure of the defects, we measure the average particle displacements $u_n = x_n-a_{111}n$ near the interstitials along the defect direction, where $x_n$ is the position of particle $n$ along the defect, and $a_{111}$ is the crystal lattice spacing along the $\left< 111 \right>$  direction. We choose $n=0$ to correspond to the particle just before the defect center and use ``standard'' boundary conditions: $u_{n=-\infty}=a_{111}$, $u_{n=\infty}=0$. We plot this displacement field for a BCC crystal with $\kappa=3.5$ and $\Gamma=2400$ in Fig. \ref{fig:crowdion}, along with the displacement field along the three other $\left< 111 \right>$ lines. Clearly, particle positions along the defect direction are strongly affected by the presence of the interstitial, while along the other $\left< 111 \right>$ directions they remain essentially unperturbed -- indicating that the defect is one-dimensional.  

A classic characteristic of a crowdion is that the defect shape can be well captured by the Frenkel-Kontorova model~\cite{kontorova1938theory,dudarev2003coherent}. This model described a one dimensional chain of particles that are connected to their neighbours via springs, and embedded in a periodic external potential.  Defects are included as missing or extra particles with respect to the number of external periodic wells. In the continuum limit, the average particle positions near a defect follow the soliton solution to the sine-Gordon equation, which has a single free parameter that captures the extent of the defect. Hence, 
to further confirm that the interstitials are realizations of crowdions  we compare our results to the soliton solution of the sine-Gordon equation (black dashed line in Fig. \ref{fig:crowdion}b), using the extension of the defect as a fit parameter.  We observe excellent agreement. Curious as to how the shape of the defect is dependent on where we are on the phase diagram, we performed the same analysis for a range of different state points with different values of $\kappa$ and $\Gamma$. Remarkably, in the area of the phase diagram close to the melting line, we see very little effect of either parameter on the structure of the crowdion. In Fig. \ref{fig:crowdion}c, we show the displacement fields along the defect axis for 15 different state points, and obtain essentially the same curve every time. This indicates that in this regime, the extension of the defect is largely independent of both the interaction strength and screening parameter. We note, however, that the crowdions do slowly get longer as $\Gamma$ is increased far beyond the melting point. This can be observed from the shape of the quenched defects, as shown in Fig. \ref{fig:defectstructure2}d. 

\section{Conclusions}

In summary, we have characterized the point defects that appear in crystals of charged colloids, an archetypical colloidal model system. Surprisingly, we found dramatic differences between the two -- fairly similar -- crystal phases FCC and BCC. As a first observation, the BCC crystals contain dramatically more vacancies, as well as more interstitials, than their FCC counterparts. One logical explanation for this is the relatively small number of nearest-neighbors in the BCC crystals, which makes it easier for particles to partially emerge from their cages and make use of the extra room opened up by a vacancy -- or adapt to the encroachment of a nearby interstitial. 

BCC not only exhibits significantly more defects, but its interstitials manifest as exotic one-dimensional defects called crowdions \cite{paneth1950mechanism}. The delocalized nature of such defects would be expected to promote fast and strongly anisotropic diffusion of the defects through the crystal \cite{han2002self,zepeda2004strongly,dudarev2003coherent,derlet2007multiscale}. In combination with the relatively large concentration of interstitials in BCC near melting, these defects are expected to strongly impact the transport properties of the crystal, including self-diffusion and the diffusion of dopants \cite{tauber2016anomalous}.

The observation of a crowdion defect in an easy-to-realize and highly tunable colloidal  system is not only important for our understanding of this system itself, but also for understanding the nature of crowdions. In atomic systems, crowdions are both rare and hard to observe directly (almost all studies are based on simulations and theory \cite{dudarev2003coherent,derlet2007multiscale, nguyen2006self, osetsky2003one, han2002self,zepeda2004strongly}). In contrast, charged colloids can be studied in real space and real time using {\it e.g.} confocal microscopy, and hence are an ideal experimental playground for studying these defects.

Surprisingly, despite decades of intense study, it appears that this fundamental colloidal model system has not yet given up all of its secrets.

\section{Acknowledgements}

The authors thank Alfons van Blaaderen, Anna Nikolaenkova, and Emanuele Boattini for many useful discussions.   L.F. acknowledges funding from the Dutch Research Council (NWO) for a Vidi grant (Grant No. VI.VIDI.192.102). B.v.d.M. acknowledges funding from the Rubicon research program with project number 019.191EN.011, which is financed by NWO.

\section{Data Availability Statement}
The data that support the findings of this study are available from the corresponding author upon reasonable request.

\section{References}

\begin{thebibliography}{52}%
\makeatletter
\providecommand \@ifxundefined [1]{%
 \@ifx{#1\undefined}
}%
\providecommand \@ifnum [1]{%
 \ifnum #1\expandafter \@firstoftwo
 \else \expandafter \@secondoftwo
 \fi
}%
\providecommand \@ifx [1]{%
 \ifx #1\expandafter \@firstoftwo
 \else \expandafter \@secondoftwo
 \fi
}%
\providecommand \natexlab [1]{#1}%
\providecommand \enquote  [1]{``#1''}%
\providecommand \bibnamefont  [1]{#1}%
\providecommand \bibfnamefont [1]{#1}%
\providecommand \citenamefont [1]{#1}%
\providecommand \href@noop [0]{\@secondoftwo}%
\providecommand \href [0]{\begingroup \@sanitize@url \@href}%
\providecommand \@href[1]{\@@startlink{#1}\@@href}%
\providecommand \@@href[1]{\endgroup#1\@@endlink}%
\providecommand \@sanitize@url [0]{\catcode `\\12\catcode `\$12\catcode
  `\&12\catcode `\#12\catcode `\^12\catcode `\_12\catcode `\%12\relax}%
\providecommand \@@startlink[1]{}%
\providecommand \@@endlink[0]{}%
\providecommand \url  [0]{\begingroup\@sanitize@url \@url }%
\providecommand \@url [1]{\endgroup\@href {#1}{\urlprefix }}%
\providecommand \urlprefix  [0]{URL }%
\providecommand \Eprint [0]{\href }%
\providecommand \doibase [0]{http://dx.doi.org/}%
\providecommand \selectlanguage [0]{\@gobble}%
\providecommand \bibinfo  [0]{\@secondoftwo}%
\providecommand \bibfield  [0]{\@secondoftwo}%
\providecommand \translation [1]{[#1]}%
\providecommand \BibitemOpen [0]{}%
\providecommand \bibitemStop [0]{}%
\providecommand \bibitemNoStop [0]{.\EOS\space}%
\providecommand \EOS [0]{\spacefactor3000\relax}%
\providecommand \BibitemShut  [1]{\csname bibitem#1\endcsname}%
\let\auto@bib@innerbib\@empty
\bibitem [{\citenamefont {Alexander}\ \emph {et~al.}(1984)\citenamefont
  {Alexander}, \citenamefont {Chaikin}, \citenamefont {Grant}, \citenamefont
  {Morales}, \citenamefont {Pincus},\ and\ \citenamefont
  {Hone}}]{alexander1984charge}%
  \BibitemOpen
  \bibfield  {author} {\bibinfo {author} {\bibfnamefont {S.}~\bibnamefont
  {Alexander}}, \bibinfo {author} {\bibfnamefont {P.~M.}\ \bibnamefont
  {Chaikin}}, \bibinfo {author} {\bibfnamefont {P.}~\bibnamefont {Grant}},
  \bibinfo {author} {\bibfnamefont {G.~J.}\ \bibnamefont {Morales}}, \bibinfo
  {author} {\bibfnamefont {P.}~\bibnamefont {Pincus}}, \ and\ \bibinfo {author}
  {\bibfnamefont {D.}~\bibnamefont {Hone}},\ }\bibfield  {title} {\enquote
  {\bibinfo {title} {Charge renormalization, osmotic pressure, and bulk modulus
  of colloidal crystals: Theory},}\ }\href@noop {} {\bibfield  {journal}
  {\bibinfo  {journal} {J. Chem. Phys.}\ }\textbf {\bibinfo {volume} {80}},\
  \bibinfo {pages} {5776--5781} (\bibinfo {year} {1984})}\BibitemShut {NoStop}%
\bibitem [{\citenamefont {Kremer}, \citenamefont {Robbins},\ and\ \citenamefont
  {Grest}(1986)}]{kremer1986phase}%
  \BibitemOpen
  \bibfield  {author} {\bibinfo {author} {\bibfnamefont {K.}~\bibnamefont
  {Kremer}}, \bibinfo {author} {\bibfnamefont {M.~O.}\ \bibnamefont {Robbins}},
  \ and\ \bibinfo {author} {\bibfnamefont {G.~S.}\ \bibnamefont {Grest}},\
  }\bibfield  {title} {\enquote {\bibinfo {title} {Phase diagram of yukawa
  systems: model for charge-stabilized colloids},}\ }\href@noop {} {\bibfield
  {journal} {\bibinfo  {journal} {Phys. Rev. Lett.}\ }\textbf {\bibinfo
  {volume} {57}},\ \bibinfo {pages} {2694} (\bibinfo {year}
  {1986})}\BibitemShut {NoStop}%
\bibitem [{\citenamefont {Robbins}, \citenamefont {Kremer},\ and\ \citenamefont
  {Grest}(1988)}]{robbins1988phase}%
  \BibitemOpen
  \bibfield  {author} {\bibinfo {author} {\bibfnamefont {M.~O.}\ \bibnamefont
  {Robbins}}, \bibinfo {author} {\bibfnamefont {K.}~\bibnamefont {Kremer}}, \
  and\ \bibinfo {author} {\bibfnamefont {G.~S.}\ \bibnamefont {Grest}},\
  }\bibfield  {title} {\enquote {\bibinfo {title} {Phase diagram and dynamics
  of yukawa systems},}\ }\href@noop {} {\bibfield  {journal} {\bibinfo
  {journal} {J. Chem. Phys.}\ }\textbf {\bibinfo {volume} {88}},\ \bibinfo
  {pages} {3286--3312} (\bibinfo {year} {1988})}\BibitemShut {NoStop}%
\bibitem [{\citenamefont {Monovoukas}\ and\ \citenamefont
  {Gast}(1989)}]{monovoukas1989experimental}%
  \BibitemOpen
  \bibfield  {author} {\bibinfo {author} {\bibfnamefont {Y.}~\bibnamefont
  {Monovoukas}}\ and\ \bibinfo {author} {\bibfnamefont {A.~P.}\ \bibnamefont
  {Gast}},\ }\bibfield  {title} {\enquote {\bibinfo {title} {The experimental
  phase diagram of charged colloidal suspensions},}\ }\href@noop {} {\bibfield
  {journal} {\bibinfo  {journal} {J. Colloid Interface. Sci.}\ }\textbf
  {\bibinfo {volume} {128}},\ \bibinfo {pages} {533--548} (\bibinfo {year}
  {1989})}\BibitemShut {NoStop}%
\bibitem [{\citenamefont {Sirota}\ \emph {et~al.}(1989)\citenamefont {Sirota},
  \citenamefont {Ou-Yang}, \citenamefont {Sinha}, \citenamefont {Chaikin},
  \citenamefont {Axe},\ and\ \citenamefont {Fujii}}]{sirota1989complete}%
  \BibitemOpen
  \bibfield  {author} {\bibinfo {author} {\bibfnamefont {E.~B.}\ \bibnamefont
  {Sirota}}, \bibinfo {author} {\bibfnamefont {H.~D.}\ \bibnamefont {Ou-Yang}},
  \bibinfo {author} {\bibfnamefont {S.~K.}\ \bibnamefont {Sinha}}, \bibinfo
  {author} {\bibfnamefont {P.~M.}\ \bibnamefont {Chaikin}}, \bibinfo {author}
  {\bibfnamefont {J.~D.}\ \bibnamefont {Axe}}, \ and\ \bibinfo {author}
  {\bibfnamefont {Y.}~\bibnamefont {Fujii}},\ }\bibfield  {title} {\enquote
  {\bibinfo {title} {Complete phase diagram of a charged colloidal system: A
  synchro-tron x-ray scattering study},}\ }\href@noop {} {\bibfield  {journal}
  {\bibinfo  {journal} {Phys. Rev. Lett.}\ }\textbf {\bibinfo {volume} {62}},\
  \bibinfo {pages} {1524} (\bibinfo {year} {1989})}\BibitemShut {NoStop}%
\bibitem [{\citenamefont {Hamaguchi}, \citenamefont {Farouki},\ and\
  \citenamefont {Dubin}(1997)}]{hamaguchi1997triple}%
  \BibitemOpen
  \bibfield  {author} {\bibinfo {author} {\bibfnamefont {S.}~\bibnamefont
  {Hamaguchi}}, \bibinfo {author} {\bibfnamefont {R.}~\bibnamefont {Farouki}},
  \ and\ \bibinfo {author} {\bibfnamefont {D.}~\bibnamefont {Dubin}},\
  }\bibfield  {title} {\enquote {\bibinfo {title} {Triple point of {Yukawa}
  systems},}\ }\href@noop {} {\bibfield  {journal} {\bibinfo  {journal} {Phys.
  Rev. E}\ }\textbf {\bibinfo {volume} {56}},\ \bibinfo {pages} {4671}
  (\bibinfo {year} {1997})}\BibitemShut {NoStop}%
\bibitem [{\citenamefont {Hynninen}\ and\ \citenamefont
  {Dijkstra}(2003)}]{hynninen2003phase}%
  \BibitemOpen
  \bibfield  {author} {\bibinfo {author} {\bibfnamefont {A.-P.}\ \bibnamefont
  {Hynninen}}\ and\ \bibinfo {author} {\bibfnamefont {M.}~\bibnamefont
  {Dijkstra}},\ }\bibfield  {title} {\enquote {\bibinfo {title} {Phase diagram
  of hard-core repulsive {Yukawa} particles with a density-dependent
  truncation: a simple model for charged colloids},}\ }\href@noop {} {\bibfield
   {journal} {\bibinfo  {journal} {J. Phys. Condens. Matter}\ }\textbf
  {\bibinfo {volume} {15}},\ \bibinfo {pages} {S3557} (\bibinfo {year}
  {2003})}\BibitemShut {NoStop}%
\bibitem [{\citenamefont {Yethiraj}\ and\ \citenamefont {van
  Blaaderen}(2003)}]{yethiraj2003colloidal}%
  \BibitemOpen
  \bibfield  {author} {\bibinfo {author} {\bibfnamefont {A.}~\bibnamefont
  {Yethiraj}}\ and\ \bibinfo {author} {\bibfnamefont {A.}~\bibnamefont {van
  Blaaderen}},\ }\bibfield  {title} {\enquote {\bibinfo {title} {A colloidal
  model system with an interaction tunable from hard sphere to soft and
  dipolar},}\ }\href@noop {} {\bibfield  {journal} {\bibinfo  {journal}
  {Nature}\ }\textbf {\bibinfo {volume} {421}},\ \bibinfo {pages} {513--517}
  (\bibinfo {year} {2003})}\BibitemShut {NoStop}%
\bibitem [{\citenamefont {Hsu}, \citenamefont {Dufresne},\ and\ \citenamefont
  {Weitz}(2005)}]{hsu2005charge}%
  \BibitemOpen
  \bibfield  {author} {\bibinfo {author} {\bibfnamefont {M.~F.}\ \bibnamefont
  {Hsu}}, \bibinfo {author} {\bibfnamefont {E.~R.}\ \bibnamefont {Dufresne}}, \
  and\ \bibinfo {author} {\bibfnamefont {D.~A.}\ \bibnamefont {Weitz}},\
  }\bibfield  {title} {\enquote {\bibinfo {title} {Charge stabilization in
  nonpolar solvents},}\ }\href@noop {} {\bibfield  {journal} {\bibinfo
  {journal} {Langmuir}\ }\textbf {\bibinfo {volume} {21}},\ \bibinfo {pages}
  {4881--4887} (\bibinfo {year} {2005})}\BibitemShut {NoStop}%
\bibitem [{\citenamefont {Royall}\ \emph {et~al.}(2006)\citenamefont {Royall},
  \citenamefont {Leunissen}, \citenamefont {Hynninen}, \citenamefont
  {Dijkstra},\ and\ \citenamefont {van Blaaderen}}]{royall2006re}%
  \BibitemOpen
  \bibfield  {author} {\bibinfo {author} {\bibfnamefont {C.~P.}\ \bibnamefont
  {Royall}}, \bibinfo {author} {\bibfnamefont {M.~E.}\ \bibnamefont
  {Leunissen}}, \bibinfo {author} {\bibfnamefont {A.-P.}\ \bibnamefont
  {Hynninen}}, \bibinfo {author} {\bibfnamefont {M.}~\bibnamefont {Dijkstra}},
  \ and\ \bibinfo {author} {\bibfnamefont {A.}~\bibnamefont {van Blaaderen}},\
  }\bibfield  {title} {\enquote {\bibinfo {title} {Re-entrant melting and
  freezing in a model system of charged colloids},}\ }\href@noop {} {\bibfield
  {journal} {\bibinfo  {journal} {J. Chem. Phys.}\ }\textbf {\bibinfo {volume}
  {124}},\ \bibinfo {pages} {244706} (\bibinfo {year} {2006})}\BibitemShut
  {NoStop}%
\bibitem [{\citenamefont {El~Masri}\ \emph {et~al.}(2011)\citenamefont
  {El~Masri}, \citenamefont {van Oostrum}, \citenamefont {Smallenburg},
  \citenamefont {Vissers}, \citenamefont {Imhof}, \citenamefont {Dijkstra},\
  and\ \citenamefont {van Blaaderen}}]{el2011measuring}%
  \BibitemOpen
  \bibfield  {author} {\bibinfo {author} {\bibfnamefont {D.}~\bibnamefont
  {El~Masri}}, \bibinfo {author} {\bibfnamefont {P.}~\bibnamefont {van
  Oostrum}}, \bibinfo {author} {\bibfnamefont {F.}~\bibnamefont {Smallenburg}},
  \bibinfo {author} {\bibfnamefont {T.}~\bibnamefont {Vissers}}, \bibinfo
  {author} {\bibfnamefont {A.}~\bibnamefont {Imhof}}, \bibinfo {author}
  {\bibfnamefont {M.}~\bibnamefont {Dijkstra}}, \ and\ \bibinfo {author}
  {\bibfnamefont {A.}~\bibnamefont {van Blaaderen}},\ }\bibfield  {title}
  {\enquote {\bibinfo {title} {Measuring colloidal forces from particle
  position deviations inside an optical trap},}\ }\href@noop {} {\bibfield
  {journal} {\bibinfo  {journal} {Soft Matter}\ }\textbf {\bibinfo {volume}
  {7}},\ \bibinfo {pages} {3462--3466} (\bibinfo {year} {2011})}\BibitemShut
  {NoStop}%
\bibitem [{\citenamefont {Smallenburg}\ \emph {et~al.}(2011)\citenamefont
  {Smallenburg}, \citenamefont {Boon}, \citenamefont {Kater}, \citenamefont
  {Dijkstra},\ and\ \citenamefont {van Roij}}]{smallenburg2011phase}%
  \BibitemOpen
  \bibfield  {author} {\bibinfo {author} {\bibfnamefont {F.}~\bibnamefont
  {Smallenburg}}, \bibinfo {author} {\bibfnamefont {N.}~\bibnamefont {Boon}},
  \bibinfo {author} {\bibfnamefont {M.}~\bibnamefont {Kater}}, \bibinfo
  {author} {\bibfnamefont {M.}~\bibnamefont {Dijkstra}}, \ and\ \bibinfo
  {author} {\bibfnamefont {R.}~\bibnamefont {van Roij}},\ }\bibfield  {title}
  {\enquote {\bibinfo {title} {Phase diagrams of colloidal spheres with a
  constant zeta-potential},}\ }\href@noop {} {\bibfield  {journal} {\bibinfo
  {journal} {J. Chem. Phys.}\ }\textbf {\bibinfo {volume} {134}},\ \bibinfo
  {pages} {074505} (\bibinfo {year} {2011})}\BibitemShut {NoStop}%
\bibitem [{\citenamefont {Kanai}\ \emph {et~al.}(2015)\citenamefont {Kanai},
  \citenamefont {Boon}, \citenamefont {Lu}, \citenamefont {Sloutskin},
  \citenamefont {Schofield}, \citenamefont {Smallenburg}, \citenamefont {van
  Roij}, \citenamefont {Dijkstra}, \citenamefont {Weitz} \emph
  {et~al.}}]{kanai2015crystallization}%
  \BibitemOpen
  \bibfield  {author} {\bibinfo {author} {\bibfnamefont {T.}~\bibnamefont
  {Kanai}}, \bibinfo {author} {\bibfnamefont {N.}~\bibnamefont {Boon}},
  \bibinfo {author} {\bibfnamefont {P.~J.}\ \bibnamefont {Lu}}, \bibinfo
  {author} {\bibfnamefont {E.}~\bibnamefont {Sloutskin}}, \bibinfo {author}
  {\bibfnamefont {A.~B.}\ \bibnamefont {Schofield}}, \bibinfo {author}
  {\bibfnamefont {F.}~\bibnamefont {Smallenburg}}, \bibinfo {author}
  {\bibfnamefont {R.}~\bibnamefont {van Roij}}, \bibinfo {author}
  {\bibfnamefont {M.}~\bibnamefont {Dijkstra}}, \bibinfo {author}
  {\bibfnamefont {D.~A.}\ \bibnamefont {Weitz}},  \emph {et~al.},\ }\bibfield
  {title} {\enquote {\bibinfo {title} {Crystallization and reentrant melting of
  charged colloids in nonpolar solvents},}\ }\href@noop {} {\bibfield
  {journal} {\bibinfo  {journal} {Phys. Rev. E}\ }\textbf {\bibinfo {volume}
  {91}},\ \bibinfo {pages} {030301} (\bibinfo {year} {2015})}\BibitemShut
  {NoStop}%
\bibitem [{\citenamefont {Arai}\ and\ \citenamefont
  {Tanaka}(2017)}]{arai2017surface}%
  \BibitemOpen
  \bibfield  {author} {\bibinfo {author} {\bibfnamefont {S.}~\bibnamefont
  {Arai}}\ and\ \bibinfo {author} {\bibfnamefont {H.}~\bibnamefont {Tanaka}},\
  }\bibfield  {title} {\enquote {\bibinfo {title} {Surface-assisted
  single-crystal formation of charged colloids},}\ }\href@noop {} {\bibfield
  {journal} {\bibinfo  {journal} {Nat. Phys.}\ }\textbf {\bibinfo {volume}
  {13}},\ \bibinfo {pages} {503--509} (\bibinfo {year} {2017})}\BibitemShut
  {NoStop}%
\bibitem [{\citenamefont {Chaudhuri}\ \emph {et~al.}(2017)\citenamefont
  {Chaudhuri}, \citenamefont {Allahyarov}, \citenamefont {L{\"o}wen},
  \citenamefont {Egelhaaf},\ and\ \citenamefont {Weitz}}]{chaudhuri2017triple}%
  \BibitemOpen
  \bibfield  {author} {\bibinfo {author} {\bibfnamefont {M.}~\bibnamefont
  {Chaudhuri}}, \bibinfo {author} {\bibfnamefont {E.}~\bibnamefont
  {Allahyarov}}, \bibinfo {author} {\bibfnamefont {H.}~\bibnamefont
  {L{\"o}wen}}, \bibinfo {author} {\bibfnamefont {S.~U.}\ \bibnamefont
  {Egelhaaf}}, \ and\ \bibinfo {author} {\bibfnamefont {D.~A.}\ \bibnamefont
  {Weitz}},\ }\bibfield  {title} {\enquote {\bibinfo {title} {Triple junction
  at the triple point resolved on the individual particle level},}\ }\href@noop
  {} {\bibfield  {journal} {\bibinfo  {journal} {Phys. Rev. Lett.}\ }\textbf
  {\bibinfo {volume} {119}},\ \bibinfo {pages} {128001} (\bibinfo {year}
  {2017})}\BibitemShut {NoStop}%
\bibitem [{\citenamefont {van Gruijthuijsen}\ \emph {et~al.}(2013)\citenamefont
  {van Gruijthuijsen}, \citenamefont {Obiols-Rabasa}, \citenamefont {Heinen},
  \citenamefont {N{\"a}gele},\ and\ \citenamefont
  {Stradner}}]{van2013sterically}%
  \BibitemOpen
  \bibfield  {author} {\bibinfo {author} {\bibfnamefont {K.}~\bibnamefont {van
  Gruijthuijsen}}, \bibinfo {author} {\bibfnamefont {M.}~\bibnamefont
  {Obiols-Rabasa}}, \bibinfo {author} {\bibfnamefont {M.}~\bibnamefont
  {Heinen}}, \bibinfo {author} {\bibfnamefont {G.}~\bibnamefont {N{\"a}gele}},
  \ and\ \bibinfo {author} {\bibfnamefont {A.}~\bibnamefont {Stradner}},\
  }\bibfield  {title} {\enquote {\bibinfo {title} {Sterically stabilized
  colloids with tunable repulsions},}\ }\href@noop {} {\bibfield  {journal}
  {\bibinfo  {journal} {Langmuir}\ }\textbf {\bibinfo {volume} {29}},\ \bibinfo
  {pages} {11199--11207} (\bibinfo {year} {2013})}\BibitemShut {NoStop}%
\bibitem [{\citenamefont {Kodger}, \citenamefont {Guerra},\ and\ \citenamefont
  {Sprakel}(2015)}]{kodger2015precise}%
  \BibitemOpen
  \bibfield  {author} {\bibinfo {author} {\bibfnamefont {T.~E.}\ \bibnamefont
  {Kodger}}, \bibinfo {author} {\bibfnamefont {R.~E.}\ \bibnamefont {Guerra}},
  \ and\ \bibinfo {author} {\bibfnamefont {J.}~\bibnamefont {Sprakel}},\
  }\bibfield  {title} {\enquote {\bibinfo {title} {Precise colloids with
  tunable interactions for confocal microscopy},}\ }\href@noop {} {\bibfield
  {journal} {\bibinfo  {journal} {Sci. Rep.}\ }\textbf {\bibinfo {volume}
  {5}},\ \bibinfo {pages} {14635} (\bibinfo {year} {2015})}\BibitemShut
  {NoStop}%
\bibitem [{\citenamefont {Yan}\ \emph {et~al.}(2005)\citenamefont {Yan},
  \citenamefont {Chen}, \citenamefont {Chua},\ and\ \citenamefont
  {Zhao}}]{yan2005incorporation}%
  \BibitemOpen
  \bibfield  {author} {\bibinfo {author} {\bibfnamefont {Q.}~\bibnamefont
  {Yan}}, \bibinfo {author} {\bibfnamefont {A.}~\bibnamefont {Chen}}, \bibinfo
  {author} {\bibfnamefont {S.~J.}\ \bibnamefont {Chua}}, \ and\ \bibinfo
  {author} {\bibfnamefont {X.~S.}\ \bibnamefont {Zhao}},\ }\bibfield  {title}
  {\enquote {\bibinfo {title} {Incorporation of point defects into
  self-assembled three-dimensional colloidal crystals},}\ }\href@noop {}
  {\bibfield  {journal} {\bibinfo  {journal} {Adv. Mater.}\ }\textbf {\bibinfo
  {volume} {17}},\ \bibinfo {pages} {2849--2853} (\bibinfo {year}
  {2005})}\BibitemShut {NoStop}%
\bibitem [{\citenamefont {Rengarajan}\ \emph {et~al.}(2005)\citenamefont
  {Rengarajan}, \citenamefont {Mittleman}, \citenamefont {Rich},\ and\
  \citenamefont {Colvin}}]{rengarajan2005effect}%
  \BibitemOpen
  \bibfield  {author} {\bibinfo {author} {\bibfnamefont {R.}~\bibnamefont
  {Rengarajan}}, \bibinfo {author} {\bibfnamefont {D.}~\bibnamefont
  {Mittleman}}, \bibinfo {author} {\bibfnamefont {C.}~\bibnamefont {Rich}}, \
  and\ \bibinfo {author} {\bibfnamefont {V.}~\bibnamefont {Colvin}},\
  }\bibfield  {title} {\enquote {\bibinfo {title} {Effect of disorder on the
  optical properties of colloidal crystals},}\ }\href@noop {} {\bibfield
  {journal} {\bibinfo  {journal} {Phys. Rev. E}\ }\textbf {\bibinfo {volume}
  {71}},\ \bibinfo {pages} {016615} (\bibinfo {year} {2005})}\BibitemShut
  {NoStop}%
\bibitem [{\citenamefont {Nelson}\ \emph {et~al.}(2011)\citenamefont {Nelson},
  \citenamefont {Dias}, \citenamefont {Bassett}, \citenamefont {Dunham},
  \citenamefont {Verma}, \citenamefont {Miyake}, \citenamefont {Wiltzius},
  \citenamefont {Rogers}, \citenamefont {Coleman}, \citenamefont {Li} \emph
  {et~al.}}]{nelson2011epitaxial}%
  \BibitemOpen
  \bibfield  {author} {\bibinfo {author} {\bibfnamefont {E.~C.}\ \bibnamefont
  {Nelson}}, \bibinfo {author} {\bibfnamefont {N.~L.}\ \bibnamefont {Dias}},
  \bibinfo {author} {\bibfnamefont {K.~P.}\ \bibnamefont {Bassett}}, \bibinfo
  {author} {\bibfnamefont {S.~N.}\ \bibnamefont {Dunham}}, \bibinfo {author}
  {\bibfnamefont {V.}~\bibnamefont {Verma}}, \bibinfo {author} {\bibfnamefont
  {M.}~\bibnamefont {Miyake}}, \bibinfo {author} {\bibfnamefont
  {P.}~\bibnamefont {Wiltzius}}, \bibinfo {author} {\bibfnamefont {J.~A.}\
  \bibnamefont {Rogers}}, \bibinfo {author} {\bibfnamefont {J.~J.}\
  \bibnamefont {Coleman}}, \bibinfo {author} {\bibfnamefont {X.}~\bibnamefont
  {Li}},  \emph {et~al.},\ }\bibfield  {title} {\enquote {\bibinfo {title}
  {Epitaxial growth of three-dimensionally architectured optoelectronic
  devices},}\ }\href@noop {} {\bibfield  {journal} {\bibinfo  {journal} {Nat.
  Mater.}\ }\textbf {\bibinfo {volume} {10}},\ \bibinfo {pages} {676--681}
  (\bibinfo {year} {2011})}\BibitemShut {NoStop}%
\bibitem [{\citenamefont {Bennett}\ and\ \citenamefont
  {Alder}(1971{\natexlab{a}})}]{bennett1971studies}%
  \BibitemOpen
  \bibfield  {author} {\bibinfo {author} {\bibfnamefont {C.}~\bibnamefont
  {Bennett}}\ and\ \bibinfo {author} {\bibfnamefont {B.}~\bibnamefont
  {Alder}},\ }\bibfield  {title} {\enquote {\bibinfo {title} {Studies in
  molecular dynamics. {IX.} {Vacancies} in hard sphere crystals},}\ }\href@noop
  {} {\bibfield  {journal} {\bibinfo  {journal} {J. Chem. Phys.}\ }\textbf
  {\bibinfo {volume} {54}},\ \bibinfo {pages} {4796--4808} (\bibinfo {year}
  {1971}{\natexlab{a}})}\BibitemShut {NoStop}%
\bibitem [{\citenamefont {Pronk}\ and\ \citenamefont
  {Frenkel}(2001)}]{pronk2001point}%
  \BibitemOpen
  \bibfield  {author} {\bibinfo {author} {\bibfnamefont {S.}~\bibnamefont
  {Pronk}}\ and\ \bibinfo {author} {\bibfnamefont {D.}~\bibnamefont
  {Frenkel}},\ }\bibfield  {title} {\enquote {\bibinfo {title} {Point defects
  in hard-sphere crystals},}\ }\href@noop {} {\bibfield  {journal} {\bibinfo
  {journal} {J. Phys. Chem. B}\ }\textbf {\bibinfo {volume} {105}},\ \bibinfo
  {pages} {6722--6727} (\bibinfo {year} {2001})}\BibitemShut {NoStop}%
\bibitem [{\citenamefont {Pronk}\ and\ \citenamefont
  {Frenkel}(2004)}]{pronk2004large}%
  \BibitemOpen
  \bibfield  {author} {\bibinfo {author} {\bibfnamefont {S.}~\bibnamefont
  {Pronk}}\ and\ \bibinfo {author} {\bibfnamefont {D.}~\bibnamefont
  {Frenkel}},\ }\bibfield  {title} {\enquote {\bibinfo {title} {Large effect of
  polydispersity on defect concentrations in colloidal crystals},}\ }\href@noop
  {} {\bibfield  {journal} {\bibinfo  {journal} {J. Chem. Phys.}\ }\textbf
  {\bibinfo {volume} {120}},\ \bibinfo {pages} {6764--6768} (\bibinfo {year}
  {2004})}\BibitemShut {NoStop}%
\bibitem [{\citenamefont {Lin}\ \emph {et~al.}(2016)\citenamefont {Lin},
  \citenamefont {Bierbaum}, \citenamefont {Schall}, \citenamefont {Sethna},\
  and\ \citenamefont {Cohen}}]{lin2016measuring}%
  \BibitemOpen
  \bibfield  {author} {\bibinfo {author} {\bibfnamefont {N.~Y.~C.}\
  \bibnamefont {Lin}}, \bibinfo {author} {\bibfnamefont {M.}~\bibnamefont
  {Bierbaum}}, \bibinfo {author} {\bibfnamefont {P.}~\bibnamefont {Schall}},
  \bibinfo {author} {\bibfnamefont {J.~P.}\ \bibnamefont {Sethna}}, \ and\
  \bibinfo {author} {\bibfnamefont {I.}~\bibnamefont {Cohen}},\ }\bibfield
  {title} {\enquote {\bibinfo {title} {Measuring nonlinear stresses generated
  by defects in 3d colloidal crystals},}\ }\href@noop {} {\bibfield  {journal}
  {\bibinfo  {journal} {Nat. Mater.}\ }\textbf {\bibinfo {volume} {15}},\
  \bibinfo {pages} {1172--1176} (\bibinfo {year} {2016})}\BibitemShut {NoStop}%
\bibitem [{\citenamefont {van~der Meer}, \citenamefont {Dijkstra},\ and\
  \citenamefont {Filion}(2017)}]{van2017diffusion}%
  \BibitemOpen
  \bibfield  {author} {\bibinfo {author} {\bibfnamefont {B.}~\bibnamefont
  {van~der Meer}}, \bibinfo {author} {\bibfnamefont {M.}~\bibnamefont
  {Dijkstra}}, \ and\ \bibinfo {author} {\bibfnamefont {L.}~\bibnamefont
  {Filion}},\ }\bibfield  {title} {\enquote {\bibinfo {title} {Diffusion and
  interactions of point defects in hard-sphere crystals},}\ }\href@noop {}
  {\bibfield  {journal} {\bibinfo  {journal} {J. Chem. Phys.}\ }\textbf
  {\bibinfo {volume} {146}},\ \bibinfo {pages} {244905} (\bibinfo {year}
  {2017})}\BibitemShut {NoStop}%
\bibitem [{\citenamefont {VanSaders}, \citenamefont {Dshemuchadse},\ and\
  \citenamefont {Glotzer}(2018)}]{vansaders2018strain}%
  \BibitemOpen
  \bibfield  {author} {\bibinfo {author} {\bibfnamefont {B.}~\bibnamefont
  {VanSaders}}, \bibinfo {author} {\bibfnamefont {J.}~\bibnamefont
  {Dshemuchadse}}, \ and\ \bibinfo {author} {\bibfnamefont {S.~C.}\
  \bibnamefont {Glotzer}},\ }\bibfield  {title} {\enquote {\bibinfo {title}
  {Strain fields in repulsive colloidal crystals},}\ }\href@noop {} {\bibfield
  {journal} {\bibinfo  {journal} {Phys. Rev. Mater.}\ }\textbf {\bibinfo
  {volume} {2}},\ \bibinfo {pages} {063604} (\bibinfo {year}
  {2018})}\BibitemShut {NoStop}%
\bibitem [{\citenamefont {Bennett}\ and\ \citenamefont
  {Alder}(1971{\natexlab{b}})}]{bennett1971persistence}%
  \BibitemOpen
  \bibfield  {author} {\bibinfo {author} {\bibfnamefont {C.}~\bibnamefont
  {Bennett}}\ and\ \bibinfo {author} {\bibfnamefont {B.}~\bibnamefont
  {Alder}},\ }\bibfield  {title} {\enquote {\bibinfo {title} {Persistence of
  vacancy motion in hard sphere crystals},}\ }\href@noop {} {\bibfield
  {journal} {\bibinfo  {journal} {J. Phys. Chem. Solids}\ }\textbf {\bibinfo
  {volume} {32}},\ \bibinfo {pages} {2111--2122} (\bibinfo {year}
  {1971}{\natexlab{b}})}\BibitemShut {NoStop}%
\bibitem [{\citenamefont {Hoogenboom}\ \emph {et~al.}(2002)\citenamefont
  {Hoogenboom}, \citenamefont {Derks}, \citenamefont {Vergeer},\ and\
  \citenamefont {van Blaaderen}}]{hoogenboom2002stacking}%
  \BibitemOpen
  \bibfield  {author} {\bibinfo {author} {\bibfnamefont {J.~P.}\ \bibnamefont
  {Hoogenboom}}, \bibinfo {author} {\bibfnamefont {D.}~\bibnamefont {Derks}},
  \bibinfo {author} {\bibfnamefont {P.}~\bibnamefont {Vergeer}}, \ and\
  \bibinfo {author} {\bibfnamefont {A.}~\bibnamefont {van Blaaderen}},\
  }\bibfield  {title} {\enquote {\bibinfo {title} {Stacking faults in colloidal
  crystals grown by sedimentation},}\ }\href@noop {} {\bibfield  {journal}
  {\bibinfo  {journal} {J. Chem. Phys.}\ }\textbf {\bibinfo {volume} {117}},\
  \bibinfo {pages} {11320--11328} (\bibinfo {year} {2002})}\BibitemShut
  {NoStop}%
\bibitem [{\citenamefont {Pronk}\ and\ \citenamefont
  {Frenkel}(1999)}]{pronk1999can}%
  \BibitemOpen
  \bibfield  {author} {\bibinfo {author} {\bibfnamefont {S.}~\bibnamefont
  {Pronk}}\ and\ \bibinfo {author} {\bibfnamefont {D.}~\bibnamefont
  {Frenkel}},\ }\bibfield  {title} {\enquote {\bibinfo {title} {Can stacking
  faults in hard-sphere crystals anneal out spontaneously?}}\ }\href@noop {}
  {\bibfield  {journal} {\bibinfo  {journal} {J. Chem. Phys}\ }\textbf
  {\bibinfo {volume} {110}},\ \bibinfo {pages} {4589--4592} (\bibinfo {year}
  {1999})}\BibitemShut {NoStop}%
\bibitem [{\citenamefont {Marechal}, \citenamefont {Hermes},\ and\
  \citenamefont {Dijkstra}(2011)}]{marechal2011stacking}%
  \BibitemOpen
  \bibfield  {author} {\bibinfo {author} {\bibfnamefont {M.}~\bibnamefont
  {Marechal}}, \bibinfo {author} {\bibfnamefont {M.}~\bibnamefont {Hermes}}, \
  and\ \bibinfo {author} {\bibfnamefont {M.}~\bibnamefont {Dijkstra}},\
  }\bibfield  {title} {\enquote {\bibinfo {title} {Stacking in sediments of
  colloidal hard spheres},}\ }\href@noop {} {\bibfield  {journal} {\bibinfo
  {journal} {J. Chem. Phys.}\ }\textbf {\bibinfo {volume} {135}},\ \bibinfo
  {pages} {034510} (\bibinfo {year} {2011})}\BibitemShut {NoStop}%
\bibitem [{\citenamefont {Pusey}\ \emph {et~al.}(1989)\citenamefont {Pusey},
  \citenamefont {Van~Megen}, \citenamefont {Bartlett}, \citenamefont
  {Ackerson}, \citenamefont {Rarity},\ and\ \citenamefont
  {Underwood}}]{pusey1989structure}%
  \BibitemOpen
  \bibfield  {author} {\bibinfo {author} {\bibfnamefont {P.~N.}\ \bibnamefont
  {Pusey}}, \bibinfo {author} {\bibfnamefont {W.}~\bibnamefont {Van~Megen}},
  \bibinfo {author} {\bibfnamefont {P.}~\bibnamefont {Bartlett}}, \bibinfo
  {author} {\bibfnamefont {B.~J.}\ \bibnamefont {Ackerson}}, \bibinfo {author}
  {\bibfnamefont {J.~G.}\ \bibnamefont {Rarity}}, \ and\ \bibinfo {author}
  {\bibfnamefont {S.~M.}\ \bibnamefont {Underwood}},\ }\bibfield  {title}
  {\enquote {\bibinfo {title} {Structure of crystals of hard colloidal
  spheres},}\ }\href@noop {} {\bibfield  {journal} {\bibinfo  {journal} {Phys.
  Rev. Lett.}\ }\textbf {\bibinfo {volume} {63}},\ \bibinfo {pages} {2753}
  (\bibinfo {year} {1989})}\BibitemShut {NoStop}%
\bibitem [{\citenamefont {van Damme}\ \emph {et~al.}(2017)\citenamefont {van
  Damme}, \citenamefont {van~der Meer}, \citenamefont {van~den Broeke},
  \citenamefont {Smallenburg},\ and\ \citenamefont {Filion}}]{van2017phase}%
  \BibitemOpen
  \bibfield  {author} {\bibinfo {author} {\bibfnamefont {R.}~\bibnamefont {van
  Damme}}, \bibinfo {author} {\bibfnamefont {B.}~\bibnamefont {van~der Meer}},
  \bibinfo {author} {\bibfnamefont {J.}~\bibnamefont {van~den Broeke}},
  \bibinfo {author} {\bibfnamefont {F.}~\bibnamefont {Smallenburg}}, \ and\
  \bibinfo {author} {\bibfnamefont {L.}~\bibnamefont {Filion}},\ }\bibfield
  {title} {\enquote {\bibinfo {title} {Phase and vacancy behaviour of hard
  “slanted” cubes},}\ }\href@noop {} {\bibfield  {journal} {\bibinfo
  {journal} {J. Chem. Phys.}\ }\textbf {\bibinfo {volume} {147}},\ \bibinfo
  {pages} {124501} (\bibinfo {year} {2017})}\BibitemShut {NoStop}%
\bibitem [{\citenamefont {van~der Meer}\ \emph {et~al.}(2018)\citenamefont
  {van~der Meer}, \citenamefont {Van~Damme}, \citenamefont {Dijkstra},
  \citenamefont {Smallenburg},\ and\ \citenamefont
  {Filion}}]{van2018revealing}%
  \BibitemOpen
  \bibfield  {author} {\bibinfo {author} {\bibfnamefont {B.}~\bibnamefont
  {van~der Meer}}, \bibinfo {author} {\bibfnamefont {R.}~\bibnamefont
  {Van~Damme}}, \bibinfo {author} {\bibfnamefont {M.}~\bibnamefont {Dijkstra}},
  \bibinfo {author} {\bibfnamefont {F.}~\bibnamefont {Smallenburg}}, \ and\
  \bibinfo {author} {\bibfnamefont {L.}~\bibnamefont {Filion}},\ }\bibfield
  {title} {\enquote {\bibinfo {title} {Revealing a vacancy analog of the
  crowdion interstitial in simple cubic crystals},}\ }\href@noop {} {\bibfield
  {journal} {\bibinfo  {journal} {Phys. Rev. Lett.}\ }\textbf {\bibinfo
  {volume} {121}},\ \bibinfo {pages} {258001} (\bibinfo {year}
  {2018})}\BibitemShut {NoStop}%
\bibitem [{\citenamefont {Smallenburg}\ \emph {et~al.}(2012)\citenamefont
  {Smallenburg}, \citenamefont {Filion}, \citenamefont {Marechal},\ and\
  \citenamefont {Dijkstra}}]{smallenburg2012vacancy}%
  \BibitemOpen
  \bibfield  {author} {\bibinfo {author} {\bibfnamefont {F.}~\bibnamefont
  {Smallenburg}}, \bibinfo {author} {\bibfnamefont {L.}~\bibnamefont {Filion}},
  \bibinfo {author} {\bibfnamefont {M.}~\bibnamefont {Marechal}}, \ and\
  \bibinfo {author} {\bibfnamefont {M.}~\bibnamefont {Dijkstra}},\ }\bibfield
  {title} {\enquote {\bibinfo {title} {Vacancy-stabilized crystalline order in
  hard cubes},}\ }\href@noop {} {\bibfield  {journal} {\bibinfo  {journal}
  {Proc. Natl. Acad. Sci. USA}\ }\textbf {\bibinfo {volume} {109}},\ \bibinfo
  {pages} {17886--17890} (\bibinfo {year} {2012})}\BibitemShut {NoStop}%
\bibitem [{\citenamefont {Paneth}(1950)}]{paneth1950mechanism}%
  \BibitemOpen
  \bibfield  {author} {\bibinfo {author} {\bibfnamefont {H.~R.}\ \bibnamefont
  {Paneth}},\ }\bibfield  {title} {\enquote {\bibinfo {title} {The mechanism of
  self-diffusion in alkali metals},}\ }\href@noop {} {\bibfield  {journal}
  {\bibinfo  {journal} {Phys. Rev.}\ }\textbf {\bibinfo {volume} {80}},\
  \bibinfo {pages} {708} (\bibinfo {year} {1950})}\BibitemShut {NoStop}%
\bibitem [{\citenamefont {Derlet}, \citenamefont {Nguyen-Manh},\ and\
  \citenamefont {Dudarev}(2007)}]{derlet2007multiscale}%
  \BibitemOpen
  \bibfield  {author} {\bibinfo {author} {\bibfnamefont {P.~M.}\ \bibnamefont
  {Derlet}}, \bibinfo {author} {\bibfnamefont {D.}~\bibnamefont {Nguyen-Manh}},
  \ and\ \bibinfo {author} {\bibfnamefont {S.~L.}\ \bibnamefont {Dudarev}},\
  }\bibfield  {title} {\enquote {\bibinfo {title} {Multiscale modeling of
  crowdion and vacancy defects in body-centered-cubic transition metals},}\
  }\href@noop {} {\bibfield  {journal} {\bibinfo  {journal} {Phys. Rev. B}\
  }\textbf {\bibinfo {volume} {76}},\ \bibinfo {pages} {054107} (\bibinfo
  {year} {2007})}\BibitemShut {NoStop}%
\bibitem [{\citenamefont {Nguyen-Manh}, \citenamefont {Horsfield},\ and\
  \citenamefont {Dudarev}(2006)}]{nguyen2006self}%
  \BibitemOpen
  \bibfield  {author} {\bibinfo {author} {\bibfnamefont {D.}~\bibnamefont
  {Nguyen-Manh}}, \bibinfo {author} {\bibfnamefont {A.~P.}\ \bibnamefont
  {Horsfield}}, \ and\ \bibinfo {author} {\bibfnamefont {S.~L.}\ \bibnamefont
  {Dudarev}},\ }\bibfield  {title} {\enquote {\bibinfo {title}
  {Self-interstitial atom defects in bcc transition metals: Group-specific
  trends},}\ }\href@noop {} {\bibfield  {journal} {\bibinfo  {journal} {Phys.
  Rev. B}\ }\textbf {\bibinfo {volume} {73}},\ \bibinfo {pages} {020101}
  (\bibinfo {year} {2006})}\BibitemShut {NoStop}%
\bibitem [{\citenamefont {Osetsky}\ \emph {et~al.}(2003)\citenamefont
  {Osetsky}, \citenamefont {Bacon}, \citenamefont {Serra}, \citenamefont
  {Singh},\ and\ \citenamefont {Golubov}}]{osetsky2003one}%
  \BibitemOpen
  \bibfield  {author} {\bibinfo {author} {\bibfnamefont {Y.~N.}\ \bibnamefont
  {Osetsky}}, \bibinfo {author} {\bibfnamefont {D.~J.}\ \bibnamefont {Bacon}},
  \bibinfo {author} {\bibfnamefont {A.}~\bibnamefont {Serra}}, \bibinfo
  {author} {\bibfnamefont {B.~N.}\ \bibnamefont {Singh}}, \ and\ \bibinfo
  {author} {\bibfnamefont {S.~I.}\ \bibnamefont {Golubov}},\ }\bibfield
  {title} {\enquote {\bibinfo {title} {One-dimensional atomic transport by
  clusters of self-interstitial atoms in iron and copper},}\ }\href@noop {}
  {\bibfield  {journal} {\bibinfo  {journal} {Philos. Mag.}\ }\textbf {\bibinfo
  {volume} {83}},\ \bibinfo {pages} {61--91} (\bibinfo {year}
  {2003})}\BibitemShut {NoStop}%
\bibitem [{\citenamefont {Han}\ \emph {et~al.}(2002)\citenamefont {Han},
  \citenamefont {Zepeda-Ruiz}, \citenamefont {Ackland}, \citenamefont {Car},\
  and\ \citenamefont {Srolovitz}}]{han2002self}%
  \BibitemOpen
  \bibfield  {author} {\bibinfo {author} {\bibfnamefont {S.}~\bibnamefont
  {Han}}, \bibinfo {author} {\bibfnamefont {L.~A.}\ \bibnamefont
  {Zepeda-Ruiz}}, \bibinfo {author} {\bibfnamefont {G.~J.}\ \bibnamefont
  {Ackland}}, \bibinfo {author} {\bibfnamefont {R.}~\bibnamefont {Car}}, \ and\
  \bibinfo {author} {\bibfnamefont {D.~J.}\ \bibnamefont {Srolovitz}},\
  }\bibfield  {title} {\enquote {\bibinfo {title} {Self-interstitials in {V}
  and {Mo}},}\ }\href@noop {} {\bibfield  {journal} {\bibinfo  {journal} {Phys.
  Rev. B}\ }\textbf {\bibinfo {volume} {66}},\ \bibinfo {pages} {220101}
  (\bibinfo {year} {2002})}\BibitemShut {NoStop}%
\bibitem [{\citenamefont {Zepeda-Ruiz}\ \emph {et~al.}(2004)\citenamefont
  {Zepeda-Ruiz}, \citenamefont {Rottler}, \citenamefont {Han}, \citenamefont
  {Ackland}, \citenamefont {Car},\ and\ \citenamefont
  {Srolovitz}}]{zepeda2004strongly}%
  \BibitemOpen
  \bibfield  {author} {\bibinfo {author} {\bibfnamefont {L.~A.}\ \bibnamefont
  {Zepeda-Ruiz}}, \bibinfo {author} {\bibfnamefont {J.}~\bibnamefont
  {Rottler}}, \bibinfo {author} {\bibfnamefont {S.}~\bibnamefont {Han}},
  \bibinfo {author} {\bibfnamefont {G.~J.}\ \bibnamefont {Ackland}}, \bibinfo
  {author} {\bibfnamefont {R.}~\bibnamefont {Car}}, \ and\ \bibinfo {author}
  {\bibfnamefont {D.~J.}\ \bibnamefont {Srolovitz}},\ }\bibfield  {title}
  {\enquote {\bibinfo {title} {Strongly non-arrhenius self-interstitial
  diffusion in vanadium},}\ }\href@noop {} {\bibfield  {journal} {\bibinfo
  {journal} {Phy. Rev. B}\ }\textbf {\bibinfo {volume} {70}},\ \bibinfo {pages}
  {060102} (\bibinfo {year} {2004})}\BibitemShut {NoStop}%
\bibitem [{\citenamefont {Kontorova}\ and\ \citenamefont
  {Frenkel}(1938)}]{kontorova1938theory}%
  \BibitemOpen
  \bibfield  {author} {\bibinfo {author} {\bibfnamefont {T.}~\bibnamefont
  {Kontorova}}\ and\ \bibinfo {author} {\bibfnamefont {J.}~\bibnamefont
  {Frenkel}},\ }\bibfield  {title} {\enquote {\bibinfo {title} {On the theory
  of plastic deformation and twinning. {II.}}}\ }\href@noop {} {\bibfield
  {journal} {\bibinfo  {journal} {Zh. Eksp. Teor. Fiz.}\ }\textbf {\bibinfo
  {volume} {8}},\ \bibinfo {pages} {1340--1348} (\bibinfo {year}
  {1938})}\BibitemShut {NoStop}%
\bibitem [{\citenamefont {Landau}, \citenamefont {Kovalev},\ and\ \citenamefont
  {Kondratyuk}(1993)}]{landau1993model}%
  \BibitemOpen
  \bibfield  {author} {\bibinfo {author} {\bibfnamefont {A.~I.}\ \bibnamefont
  {Landau}}, \bibinfo {author} {\bibfnamefont {A.~S.}\ \bibnamefont {Kovalev}},
  \ and\ \bibinfo {author} {\bibfnamefont {A.~D.}\ \bibnamefont {Kondratyuk}},\
  }\bibfield  {title} {\enquote {\bibinfo {title} {Model of interacting atomic
  chains and its application to the description of the crowdion in an
  anisotropic crystal},}\ }\href@noop {} {\bibfield  {journal} {\bibinfo
  {journal} {Phys. Status Solidi B}\ }\textbf {\bibinfo {volume} {179}},\
  \bibinfo {pages} {373--381} (\bibinfo {year} {1993})}\BibitemShut {NoStop}%
\bibitem [{\citenamefont {Kovalev}\ \emph {et~al.}(1993)\citenamefont
  {Kovalev}, \citenamefont {Kondratyuk}, \citenamefont {Kosevich},\ and\
  \citenamefont {Landau}}]{kovalev1993theoretical}%
  \BibitemOpen
  \bibfield  {author} {\bibinfo {author} {\bibfnamefont {A.~S.}\ \bibnamefont
  {Kovalev}}, \bibinfo {author} {\bibfnamefont {A.~D.}\ \bibnamefont
  {Kondratyuk}}, \bibinfo {author} {\bibfnamefont {A.~M.}\ \bibnamefont
  {Kosevich}}, \ and\ \bibinfo {author} {\bibfnamefont {A.~I.}\ \bibnamefont
  {Landau}},\ }\bibfield  {title} {\enquote {\bibinfo {title} {Theoretical
  description of the crowdion in an anisotropic crystal based on the
  {Frenkel-Kontorova} model including and elastic three-dimensional medium},}\
  }\href@noop {} {\bibfield  {journal} {\bibinfo  {journal} {Phys. Status
  Solidi B}\ }\textbf {\bibinfo {volume} {177}},\ \bibinfo {pages} {117--127}
  (\bibinfo {year} {1993})}\BibitemShut {NoStop}%
\bibitem [{\citenamefont {Braun}\ and\ \citenamefont
  {Kivshar}(1998)}]{braun1998nonlinear}%
  \BibitemOpen
  \bibfield  {author} {\bibinfo {author} {\bibfnamefont {O.~M.}\ \bibnamefont
  {Braun}}\ and\ \bibinfo {author} {\bibfnamefont {Y.~S.}\ \bibnamefont
  {Kivshar}},\ }\bibfield  {title} {\enquote {\bibinfo {title} {Nonlinear
  dynamics of the {Frenkel--Kontorova} model},}\ }\href@noop {} {\bibfield
  {journal} {\bibinfo  {journal} {Phys. Rep.}\ }\textbf {\bibinfo {volume}
  {306}},\ \bibinfo {pages} {1--108} (\bibinfo {year} {1998})}\BibitemShut
  {NoStop}%
\bibitem [{\citenamefont {Dudarev}(2003)}]{dudarev2003coherent}%
  \BibitemOpen
  \bibfield  {author} {\bibinfo {author} {\bibfnamefont {S.}~\bibnamefont
  {Dudarev}},\ }\bibfield  {title} {\enquote {\bibinfo {title} {Coherent motion
  of interstitial defects in a crystalline material},}\ }\href@noop {}
  {\bibfield  {journal} {\bibinfo  {journal} {Philos. Mag.}\ }\textbf {\bibinfo
  {volume} {83}},\ \bibinfo {pages} {3577--3597} (\bibinfo {year}
  {2003})}\BibitemShut {NoStop}%
\bibitem [{\citenamefont {Fitzgerald}\ and\ \citenamefont
  {Nguyen-Manh}(2008)}]{fitzgerald2008peierls}%
  \BibitemOpen
  \bibfield  {author} {\bibinfo {author} {\bibfnamefont {S.~P.}\ \bibnamefont
  {Fitzgerald}}\ and\ \bibinfo {author} {\bibfnamefont {D.}~\bibnamefont
  {Nguyen-Manh}},\ }\bibfield  {title} {\enquote {\bibinfo {title} {Peierls
  potential for crowdions in the bcc transition metals},}\ }\href@noop {}
  {\bibfield  {journal} {\bibinfo  {journal} {Phys. Rev. Lett.}\ }\textbf
  {\bibinfo {volume} {101}},\ \bibinfo {pages} {115504} (\bibinfo {year}
  {2008})}\BibitemShut {NoStop}%
\bibitem [{\citenamefont {Ivlev}\ \emph {et~al.}(2012)\citenamefont {Ivlev},
  \citenamefont {Morfill}, \citenamefont {Lowen},\ and\ \citenamefont
  {Royall}}]{ivlev2012complex}%
  \BibitemOpen
  \bibfield  {author} {\bibinfo {author} {\bibfnamefont {A.}~\bibnamefont
  {Ivlev}}, \bibinfo {author} {\bibfnamefont {G.}~\bibnamefont {Morfill}},
  \bibinfo {author} {\bibfnamefont {H.}~\bibnamefont {Lowen}}, \ and\ \bibinfo
  {author} {\bibfnamefont {C.~P.}\ \bibnamefont {Royall}},\ }\href@noop {}
  {\emph {\bibinfo {title} {Complex plasmas and colloidal dispersions:
  particle-resolved studies of classical liquids and solids}}},\ Vol.~\bibinfo
  {volume} {5}\ (\bibinfo  {publisher} {World Scientific Publishing Company},\
  \bibinfo {year} {2012})\BibitemShut {NoStop}%
\bibitem [{\citenamefont {Frenkel}\ and\ \citenamefont
  {Smit}(2001)}]{frenkel2001understanding}%
  \BibitemOpen
  \bibfield  {author} {\bibinfo {author} {\bibfnamefont {D.}~\bibnamefont
  {Frenkel}}\ and\ \bibinfo {author} {\bibfnamefont {B.}~\bibnamefont {Smit}},\
  }\href@noop {} {\emph {\bibinfo {title} {Understanding molecular simulation:
  from algorithms to applications}}},\ Vol.~\bibinfo {volume} {1}\ (\bibinfo
  {publisher} {Elsevier},\ \bibinfo {year} {2001})\BibitemShut {NoStop}%
\bibitem [{\citenamefont {Frenkel}\ and\ \citenamefont
  {Ladd}(1984)}]{frenkel1984new}%
  \BibitemOpen
  \bibfield  {author} {\bibinfo {author} {\bibfnamefont {D.}~\bibnamefont
  {Frenkel}}\ and\ \bibinfo {author} {\bibfnamefont {A.~J.~C.}\ \bibnamefont
  {Ladd}},\ }\bibfield  {title} {\enquote {\bibinfo {title} {New {M}onte
  {C}arlo method to compute the free energy of arbitrary solids. {A}pplication
  to the fcc and hcp phases of hard spheres},}\ }\href@noop {} {\bibfield
  {journal} {\bibinfo  {journal} {J. Chem. Phys.}\ }\textbf {\bibinfo {volume}
  {81}},\ \bibinfo {pages} {3188--3193} (\bibinfo {year} {1984})}\BibitemShut
  {NoStop}%
\bibitem [{Note1()}]{Note1}%
  \BibitemOpen
  \bibinfo {note} {Note that we have taken the phase boundaries from Ref.
  \protect \rev@citealpnum {hamaguchi1997triple}, and have ensured that the
  crystal phases do not melt in our simulations.}\BibitemShut {Stop}%
\bibitem [{\citenamefont {van~der Meer}\ \emph {et~al.}(2020)\citenamefont
  {van~der Meer}, \citenamefont {Smallenburg}, \citenamefont {Dijkstra},\ and\
  \citenamefont {Filion}}]{van2020high}%
  \BibitemOpen
  \bibfield  {author} {\bibinfo {author} {\bibfnamefont {B.}~\bibnamefont
  {van~der Meer}}, \bibinfo {author} {\bibfnamefont {F.}~\bibnamefont
  {Smallenburg}}, \bibinfo {author} {\bibfnamefont {M.}~\bibnamefont
  {Dijkstra}}, \ and\ \bibinfo {author} {\bibfnamefont {L.}~\bibnamefont
  {Filion}},\ }\bibfield  {title} {\enquote {\bibinfo {title} {High antisite
  defect concentrations in hard-sphere colloidal laves phases},}\ }\href@noop
  {} {\bibfield  {journal} {\bibinfo  {journal} {Soft Matter}\ } (\bibinfo
  {year} {2020})}\BibitemShut {NoStop}%
\bibitem [{\citenamefont {Tauber}, \citenamefont {Higler},\ and\ \citenamefont
  {Sprakel}(2016)}]{tauber2016anomalous}%
  \BibitemOpen
  \bibfield  {author} {\bibinfo {author} {\bibfnamefont {J.}~\bibnamefont
  {Tauber}}, \bibinfo {author} {\bibfnamefont {R.}~\bibnamefont {Higler}}, \
  and\ \bibinfo {author} {\bibfnamefont {J.}~\bibnamefont {Sprakel}},\
  }\bibfield  {title} {\enquote {\bibinfo {title} {Anomalous dynamics of
  interstitial dopants in soft crystals},}\ }\href@noop {} {\bibfield
  {journal} {\bibinfo  {journal} {Proc. Natl. Acad. Sci. USA}\ }\textbf
  {\bibinfo {volume} {113}},\ \bibinfo {pages} {13660--13665} (\bibinfo {year}
  {2016})}\BibitemShut {NoStop}%
\end{thebibliography}

%

\end{document}